\documentclass[
  journal=pasa,
  manuscript=Research-Article,
  year=2025,
  volume=42,
]{cup-journal}

\usepackage{amsmath}
\usepackage{subfigure}
\usepackage{caption, subcaption}
\usepackage[nopatch]{microtype}
\usepackage{booktabs}
\usepackage{placeins}
\usepackage{graphicx, float}
\usepackage{orcidlink}
\usepackage{url}
\hypersetup{hidelinks}

\title{TASSIE: a TASmanian Search for Inclined Exoplanets}

\author{T. Plunkett \orcidlink{0009-0003-5810-1314}}
\affiliation{Greenhill Observatory, School of Natural Sciences, University of Tasmania, Private Bag 37, Hobart, TAS 7001 Australia}
\email[T. Plunkett]{thomas.plunkett@utas.edu.au }

\author{A. A. Cole \orcidlink{0000-0003-0303-3855}}
\affiliation{Greenhill Observatory, School of Natural Sciences, University of Tasmania, Private Bag 37, Hobart, TAS 7001 Australia}

\author{J. P. Beaulieu \orcidlink{0000-0003-0014-3354}}
\affiliation{Greenhill Observatory, School of Natural Sciences, University of Tasmania, Private Bag 37, Hobart, TAS 7001 Australia}
\alsoaffiliation{Sorbonne Universit\'e, CNRS, Institut d'Astrophysique de Paris, IAP, F-75014, Paris, France}

\author{K. Siellez}
\affiliation{Greenhill Observatory, School of Natural Sciences, University of Tasmania, Private Bag 37, Hobart, TAS 7001 Australia}

\author{B. Emptage}
\affiliation{Greenhill Observatory, School of Natural Sciences, University of Tasmania, Private Bag 37, Hobart, TAS 7001 Australia}

\author{K. Auchettl \orcidlink{0000-0002-4449-9152}}
\affiliation{School of Physics, University of Melbourne, Victoria, Australia}
\alsoaffiliation{Department of Astronomy and Astrophysics, University of California, Santa Cruz, CA 95064, USA}

\author{J. W. Blackman \orcidlink{0000-0001-5860-1157}}
\affiliation{Physikalisches Institut, Universit\"{a}t Bern, Gessellschaftsstrasse 6, CH-3012 Bern, Switzerland}

\author{Natalia E. Rektsini \orcidlink{0000-0002-1530-4870}}
\affiliation{Greenhill Observatory, School of Natural Sciences, University of Tasmania, Private Bag 37, Hobart, TAS 7001 Australia}

\alsoaffiliation{Sorbonne Universit\'e, CNRS, Institut d'Astrophysique de Paris, IAP, F-75014, Paris, France}

 %% First letter not capped

\begin{document}
\newcommand{\gaia}{\textit{Gaia\ }}

\begin{abstract}
{We present the first results of a pilot `TASmanian Search for Inclined Exoplanets' (TASSIE) program. This includes observations and analysis of five short-period exoplanet candidates using data from TESS and the Harlingten 50 cm telescope at the Greenhill Observatory. We describe the instrumentation, data reduction process and target selection strategy for the program. We utilise archival multi-band photometry and new mid-resolution spectra to determine stellar parameters for five TESS Objects of Interest (TOIs). We then perform a statistical validation to rule out false positives, before moving on to a joint transit analysis of the remaining systems. We find that TOI3070, TOI3124 and TOI4266 are likely non-planetary signals, which we attribute to either short-period binary stars on grazing orbits or stellar spots. For TOI3097, we find a hot sub-Jovian to Jovian size planet ($R_{3097Ab}$ = 0.89 $\pm$ 0.04 $R_{J}$, $P_{3097Ab}$ = 1.368386 $\pm$ 0.000006 days) orbiting the primary K dwarf star in a wide binary system. This system shows indications of low metallicity ([Fe/H] $\approx$ -1), making it an unlikely host for a giant planet. For TOI3163, we find a Jovian-size companion on a circular orbit around a late F dwarf star, with $R_{3163b} = 1.42 \pm 0.05 \, R_{J}$ and $P_{3163b} = 3.074966 \pm 0.000022$ days. In future, we aim to validate further southern giant planet candidates with a particular focus on those residing in the sub-Jovian desert/savanna. }
\end{abstract}

\section{Introduction}
Giant planets orbiting close to their host stars are an uncommon product of the planet formation process. Despite their inherent rarity, the strong observational biases of the transit and radial velocity methods have allowed for the discovery and characterisation of hundreds of such close-in, giant planets \citep{WinnFab2015}. This sample has revealed multiple trends, including the correlations of giant planet occurrence with stellar mass and metallicity \citep[e.g.,][]{Johnson2010}, a lack of close companions to hot giants \citep[e.g.,][]{huang2016} and radius inflation from stellar irradiation \citep[e.g.,][]{weiss}. These facts have provided tantalising hints into the formation and evolution mechanisms of hot, giant planets \citep{DawsonJohnson2018}.
\newline\newline
Another major discovery from Kepler \citep{kepler} was a dearth of sub-Jovian planets below periods of roughly 3 days, known as the `sub-Jovian' or `Neptunian' desert \citep{subjovian, Mazeh2016}. This was soon connected with evolutionary processes, such as photoevaporation, tidal disruption and possibly migration scenarios \citep[e.g.,][]{desert}. In recent years, planets have begun to be discovered in the desert, converting the region into a so-called `savanna' \citep[for examples, see][]{savanna}. These are excellent targets for not only understanding the processes shaping the desert/savanna, but will also provide insights into giant planet formation more broadly. Furthermore, they are accessible for studies with small telescopes due to their large transit depths.  
\newline\newline 
Since the launch of the Transiting Exoplanet Survey Satellite \citep[TESS,][]{tess}, thousands of new exoplanet candidates have been reported. A major advantage of TESS over previous surveys is its nearly all-sky coverage, enabling studies of the exoplanet population across the galaxy. Modern programs are now using TESS data to investigate specific sub-populations of giant planets. For example, the `Searching for Giant Exoplanets around M-dwarf Stars' survey \citep[GEMS,][]{GEMSI} and `M dwarfs Accompanied by close-iN Giant Orbiters with SPECULOOS' project \citep[MANGOS,][]{mangos} are aiming to study the occurrence rates of giant planets around low mass stars to test classical planet formation scenarios, such as core accretion \citep[e.g.,][]{Lissauer} and gravitational instability \citep{boss1997}. This was a difficult task in the Kepler era, due to the limited sky coverage resulting in a small sample of M-dwarf hosts and, therefore, incomplete statistics. Another example is the `TESS Hunt for Young and Maturing Exoplanets' \citep[THYME,][]{thyme}. This program is searching for young transiting planets in stellar associations in order to understand evolutionary processes, including for gas giants. 
\newline\newline
With this context, we introduce the `TASmanian Search for Inclined Exoplanets' (TASSIE) follow-up program. Based at the University of Tasmania Greenhill Observatory (UTGO), we aim to study giant transiting planets in the southern sky using photometric and spectroscopic measurements. In particular, we wish to contribute to the understanding of the sub-Jovian desert/savanna through the search for more elusive planets in this region. By leveraging our own telescopic infrastructure, we hope to ensure a consistent approach to photometric data collection and analysis. A side benefit of our program will be the addition of another facility for follow-up of southern planet candidates from TESS and future space-based transit surveys. This will be useful for ruling out false positives scenarios, such as eclipsing binaries (EBs), along with refining orbital ephemerides to allow for successful observations with time-limited facilities.
\newline\newline 
In this paper, we present the first results of the TASSIE follow-up program, along with the photometric characterisation of the new Harlingten 50 cm telescope at UTGO. The paper is organised as follows. In Section 2, we introduce the instrumentation, data reduction process and target selection strategy for TASSIE. In Section 3, we present our first results with observations of five short-period TESS TOIs. In Section 4, we begin our analysis by first deriving the stellar parameters using photometric and spectroscopic measurements. We then combine astrometry, archival imaging and statistical tests with TRICERATOPS \citep{Triceratops} to rule out false positive scenarios. We then move to joint transit modelling using data from TESS and the Harlingten 50 cm. In Section 5, we discuss searching for transit timing variations and the photometric performance of our telescope. Finally, we present our conclusions in Section 6.

\section{TASSIE Follow-Up Program}
\subsection{Instrumentation - The Harlingten 50 cm Telescope}
\begin{figure}[h!]
	\centering
	\includegraphics[width=0.73\linewidth]
    {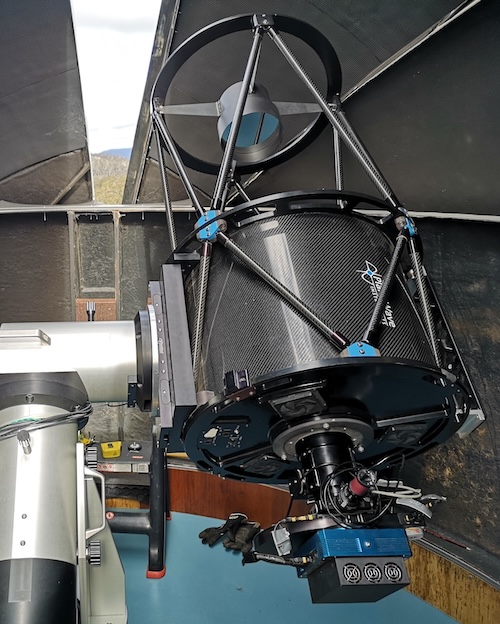}
	\caption{The Harlingten 50 cm from inside the dome, displaying the camera and filter wheel attached.}
	\label{fig:50cminside01}
\end{figure}

The Harlingten 50 cm telescope (H50, see Figure \ref{fig:50cminside01}) is a Planewave 20-inch (508 mm) of corrected Dall-Kirkham design, with a 191 mm secondary mirror. It is equipped with two corrector lenses, producing a 3454 mm focal length. The telescope is mounted onto an Alcor Systems Nova 120 direct-drive  mount. For guiding, we utilise a 648x486 SBIG ST-i autoguiding CCD with the PhD2 Guiding software\footnote{https://openphdguiding.org}. For the imaging camera, we are using an Apogee Alta U42 CCD. With 2048x2048 13$\mu$m pixels and a pixel scale of 0.8$^{\prime\prime}$/pixel, the system achieves a 27.2 sq.\ arcminute field-of-view (FOV). Accompanying this camera is an Optec TCF-S focuser (with a step resolution of 2.2 $\mu$m) and an Apogee FW50-10S filter wheel. We have Bessell B \& V and SDSS g$^{\prime}$, r$^{\prime}$, i$^{\prime}$ broadband filters, along with narrowband H$\alpha$, H$\beta$, OIII \& SII. 

\subsubsection{Noise Characteristics}
To determine the noise characteristics of the current CCD camera, we employ the methods described in \citet{Howell}. We measure the gain and read noise using
sets of short exposure and high signal-to-noise ratio (SNR) twilight flat frames and a collection of bias
frames. To measure the typical dark current, we use multiple sets of dark frames at different
exposure times. After creating a master bias frame and subtracting this from each set
of dark frames, we calculate mean and standard deviation in each set of exposures. We then convert this to electrons using the previously calculated gain. The dark current is estimated by performing a
linear regression of the mean electron count against exposure time. The measured values with the CCD cooled to the typical operational temperature (-30 $^o$C) are
shown in Table \ref{table:50cmdetails}. Empirically, we have found a magnitude limit in 1-hour integration of $r' = 20.08 \pm 0.19$ for a $SNR > 10$ in dark time. 

\begin{table}[h!]
	\begin{center}
		\begin{tabular}{|p{3cm}|p{3cm}|} 
			\toprule
			\headrow \multicolumn{2}{|c|}{Telescope (PlaneWave CDK20)}\\
			\midrule 
			Primary Mirror [mm] & 508 \\
			Focal Length [mm] & 3454 \\
			Pixel Scale [$\prime\prime$/pix] & 0.8 \\
            FOV [sq. arcminutes] & 27.2 \\
            \midrule
            \headrow \multicolumn{2}{|c|}{Mount (Alcor Nova 120)} \\
            \midrule
            RMS Guiding Error [$\prime\prime$] & 0.70 $\pm$ 0.01\\
            Tracking Drift [$\prime\prime$/hour] & West: 0.52 $\pm$ 0.04 \newline South: 1.61 $\pm$ 0.13 \\
		    \midrule
            \headrow \multicolumn{2}{|c|}{Filters} \\
            \midrule 
			Broadband Filters &  Bessell B \& V \newline SDSS g', r' \& i' \\
			Narrowband Filters & H$\alpha$, H$\beta$, OIII and SII \\
            \midrule 
            \headrow \multicolumn{2}{|c|}{Camera (Apogee ALTA U42)}\\
            \midrule 
		    Dark Current [e$^-$/pix/sec] & 0.12 $\pm$ 0.01\\
		    Read noise [e$^-$/pix] &  7.10 $\pm$ 0.10 \\
		    Gain [e$^-$/ADU] & 1.28 $\pm$ 0.02\\
            Mag. Limit (SNR > 10) & $r^{\prime}$ = 20.08 $\pm$ 0.19 \\
			\bottomrule
		\end{tabular}
		\caption{Telescope and Imaging Camera Information for the H50}
		\label{table:50cmdetails}
	\end{center}
\end{table}

\newpage

\subsection{Data Reduction and Photometry}
We have developed a predominately Python-based reduction pipeline, utilising the Prose package \citep{Garcia_2022}. An auto-reduction process performs daily checks for new data. After organisation and separation of science and calibration data, master calibration frames are produced by either a sigma-clipped mean (for bias and dark frames) or using a simple median (for flats). The science images are then reduced using standard bias, dark and flat-field corrections with the closest available calibration frames by date. Bad pixel maps are also built using the darks to help remove hot and dead pixels. The reduced science images are then passed to a locally installed copy of Astrometry.net \citep{Lang_2010} to determine the World Coordinate System (WCS) information.
\newline\newline 
For transit data, the most common tool for photometric analysis is differential aperture photometry. We implement this once again with Prose, which chooses optimal comparison stars using the algorithm outlined in Broeg et al. (2005). In summary, differential light curves are extracted for all detected sources and assigned a weighting depending upon the amount of variance. Then comparison stars are recursively rejected until only the least varying stars remain. It should be noted that this algorithm does not take into account the colour of the target, which may cause systematics in the data due to differential refraction effects. Thus, we have also implemented star selection based on closest colour and magnitude through queries to the \gaia DR2 catolog \citep{Brown_2018}.

\subsection{Target Selection Strategy}
The main goal of TASSIE is to validate giant planets, with a particular emphasis upon those residing within the sub-Jovian desert/savanna. We therefore focus on candidates with periods below $\approx$ 3 days. As a small telescope program, we are limited to relatively bright targets. Combining results from our Exposure Time Calculator (ETC)\footnote{\url{https://github.com/tjplunkett/UTGO_ObsTools}} and observations of confirmed planets, we have found that transit depths greater than 5 parts per thousand (ppt) are required for a reasonable transit SNR to be obtained with our instrument. This value is typical of other sub-metre ground-based telescopes \citep[e.g.,][]{trappist_ppt, Minerva}. This limit is adequate for detection of planets ranging from sub-Saturns to Hot Jupiters orbiting FGK stars, with the possibility to reach sub-Neptunes around M dwarf stars.
\newline\newline 
We impose a condition of exposure times below 200 seconds (with less than 120 seconds being preferred) to improve upon the TESS cadence. This allows us to search for transit timing and transit depth variations \citep[TTVs and TDVs, see][]{agolfabttvs} with greater precision. We choose to use the SDSS r' filter primarily, as it has peak quantum efficiency (QE = 0.92) for our camera. It also reduces atmospheric effects on the photometry (such as scintillation/dispersion) compared to bluer filters. This leads to a magnitude limit of r < 15 mag (for 1\% photometry), with brighter targets amenable to radial velocity follow-up being prioritised. We discuss the experimental photometric precision achieved in Section 5.3, based upon the results reported in the following sections. 

\section{Observations}
Observations for TASSIE began in the summer of 2023/2024. The targets were alerted through a custom Python script that downloads the list of TOIs each week from the Exoplanet Archive\footnote{\url{https://exoplanetarchive.ipac.caltech.edu}}. The Astroplan package \citep{astroplan} is then used to calculate the expected transit times for the next 10 epochs for any system observable from Tasmania. If any transits occur during local nighttime hours and pass our observing criteria (T < 15, $\delta$ > 5 ppt), then they are added to a target list. We show details of each observing session and target, including the coordinates, filters, exposure time, median FWHM and number of images in Table \ref{table:tassielog}. Note that the H50 telescope was not fully collimated during this period, leading to below-average FWHMs. We download all TESS target pixel files (TPF) and Science Processing Operations Center (SPOC) light curves from Mikulski Archive for Space Telescopes using the Lightkurve package \citep{Lightkurve}.

\subsection{Photometry}
\begin{table*}[ht!]
	\begin{center}
		\begin{tabular}{|p{1.5cm}|p{2.5cm}|p{1.5cm}|p{1.0cm}|p{1cm}|p{1.5cm}|p{1.25cm}|p{1.5cm}|}
			\toprule
			\headrow \textbf{ID} & \textbf{Coord. [J2000]}&\textbf{Dates [UTC]} & \textbf{Filter} & \textbf{Exp. Time [s]} & \textbf{Median FWHM ["]} & \textbf{No. of Images} & \textbf{Airmass} \\
			\midrule 
			\textbf{TOI3070} & 11:33:35.8 -56:42:08.7&16/01/24 \newline 12/02/24  & r' \newline r' & 200 \newline 200 & 4.9 \newline 3.5  &  41 \newline 24 & 1.6 - 1.2 \newline 1.4 - 1.2 \\
			\midrule 
			\textbf{TOI3097} & 11:32:17.9 -47:29:47.0 & 15/03/24  & r' & 90 & 2.7  & 91 & 1.3 - 1.0 \\
			\midrule
			\textbf{TOI3124} & 11:52:35.5 -77:08:20.4 & 17/03/24 \newline 01/07/24  & r' \newline V & 120 \newline 120 & 3.2 \newline 2.4 & 67 \newline 70  & 1.3 - 1.2 \newline 1.3 - 1.4 \\
			\midrule 
			\textbf{TOI3163} & 12:44:21.3 -55:00:47.9 &  10/02/24 & r' &  120 & 3.3  &  80 & 1.8 - 1.1 \\
			\midrule  
			\textbf{TOI4266} & 09:43:06.9 -58:30:38.8 & 15/01/24  & r' &  150 &  3.6 & 56 & 1.3 - 1.1 \\
			\bottomrule 
		\end{tabular}
		\caption{Observation logs for TASSIE targets using the H50 telescope.}
		\label{table:tassielog}
	\end{center}
\end{table*}

\subsubsection{TIC 452049244 (TOI3070)}
TIC 452049244 is a V $\approx$ 14.3 mag star in the constellation of Centaurus. TIC 452049244 was observed by TESS in Sectors 10 (26/03/19 - 22/04/19), 37 (02/04/21 - 28/04/21) and 64 (10/03/2023 - 06/04/2023). TOI3070 was originally alerted as a TOI at 04/06/2021 01:36:30 UTC using the Faint Star Search pipeline \citep[see][]{faintstar}. A clear transit-like signal is visible with an approximate period of 0.66 days, determined using a simple Box-Least Squares search \citep{BLS}. 
\newline\newline 
TOI3070 was observed with the H50 telescope on two nights. Beginning on the evening of the 16th of January 2024, 41 images were taken with an exposure time of 200 seconds in the SDSS r' filter. Whilst the sky was clear and the moon illumination low (roughly 25 \%), conditions were non-optimal with relatively strong winds resulting in a poor median FWHM of $4.9 ^{\prime\prime}$. The observations covered ingress and egress, but unfortunately captured limited out-of-transit baseline. On the 12th of February 2024, TOI3070 was re-observed with the same filter and camera settings. In this epoch the conditions were marginally better, with lower winds and clear, moonless skies. This lead to an improved median FWHM of $3.5 ^{\prime\prime}$. 24 images were obtained, covering the baseline and ingress. However, observations were interrupted due to high humidity forcing the closure of the dome early. 

\subsubsection{TIC 163262555 (TOI3097)}
TIC 163262555 is a V $\approx$ 13.8 mag star residing in the constellation of Centaurus. It was observed by TESS in Sectors 10 (26/03/19 - 22/04/19), 37 (02/04/21 - 28/04/21), 63 (10/03/2023 - 06/04/2023) and 64 (06/04/2023 - 04/05/2023). TOI3097 was originally alerted as a TOI on 04/06/2021 01:36:34 UTC from the Faint Star Search. This target is known to have a binary companion within $1.5^{\prime\prime}$ of the primary target \citep{Companions}, corresponding to a separation of 576 au. A possible transit signal is visible in the SPOC light curves, with an approximate period of 1.368 days. The periodogram and TPF are shown in the Appendix. 
\newline\newline
The TOI3097 system was observed with the H50 on the 15th of March 2024 using the SDSS r' filter. A total of 91 images were obtained, each with exposure time of 90 seconds. This covered the baseline prior to ingress through to end of egress. Sky conditions were clear and moonless. Low winds allowed for the best FWHM obtained during the summer season, with the median for the night at $2.7^{\prime\prime}$.

\subsubsection{TIC 454918388 (TOI3124)}
TIC 454918388 is a V $\approx$ 13.3 mag star in the constellation of Chamaeleon. TIC 454918388 was observed by TESS on multiple occasions, with the most recent in Sectors 64 (06/04/2023 - 04/05/2023), 65 (04/05/2023 - 02/06/2023) and 66 (02/06/2023 - 01/07/2023). TOI3124 was alerted as a TOI on 04/06/2021 13:54:19 UTC via the Faint Star Search. Inspection of the SPOC light curve shows moderate flux variations, indicating that this is a variable/active star. To determine if a planetary signal is hidden within this variability, we apply a Savitzky-Golay filter to the data \citep{savitzky}. Performing the usual BLS search on this detrended light curve gives the periodogram shown in the bottom right of Figure \ref{fig:tpflcprdmain} (a). The planetary periodogram indicates a period of $\approx$ 2.255 days, which aligns with the stellar rotation period of 2.32 days (discussed in Section 4.1.4).
\newline\newline
The H50 first observed TOI3124 on 17/03/2024 with the SDSS r' filter. The exposure times were 120 seconds, with a total of 67 images obtained. The conditions were clear and moonless with low winds to begin with, before increasing later in the night.  This resulted in a median FWHM of $3.2^{\prime\prime}$ for the observing session. We re-observed TOI3124 on 01/07/2024, this time in the Bessell V filter with exposures of 120 seconds. The sky conditions were clear and moonless, but the humidity was high. The median FWHM for this session was $3.1^{\prime\prime}$.

\subsubsection{TIC 425316308 (TOI3163)}
\begin{figure*}[h!]
    \begin{subfigure}[]
        \centering
	   \includegraphics[width = 0.475\linewidth, height = 0.45\linewidth]{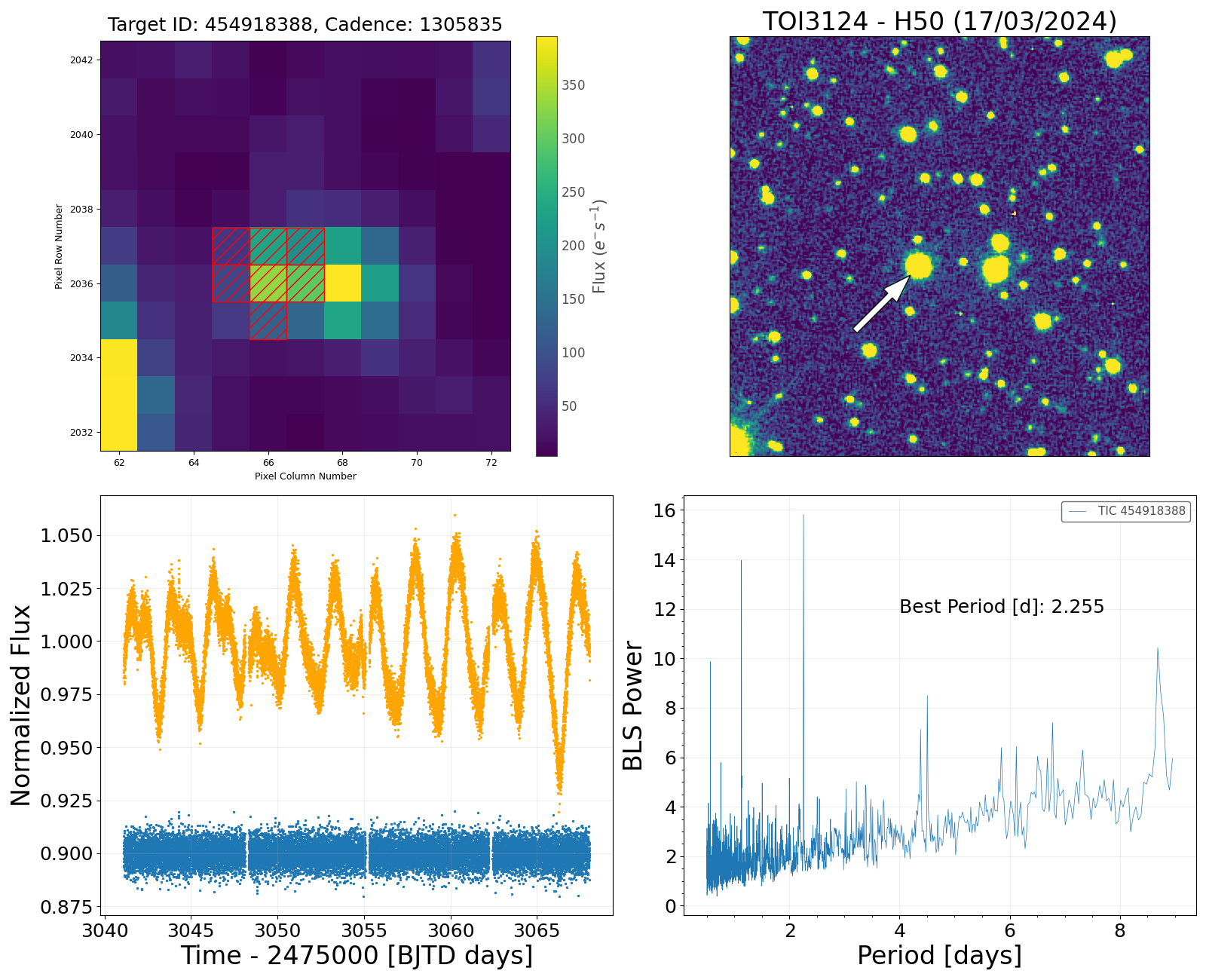}
        %\caption{}
        \label{fig:sptlcprd3124}
    \end{subfigure}
    ~
    \begin{subfigure}[]
	   \centering
	   \includegraphics[width = 0.475\linewidth, height = 0.45\linewidth]{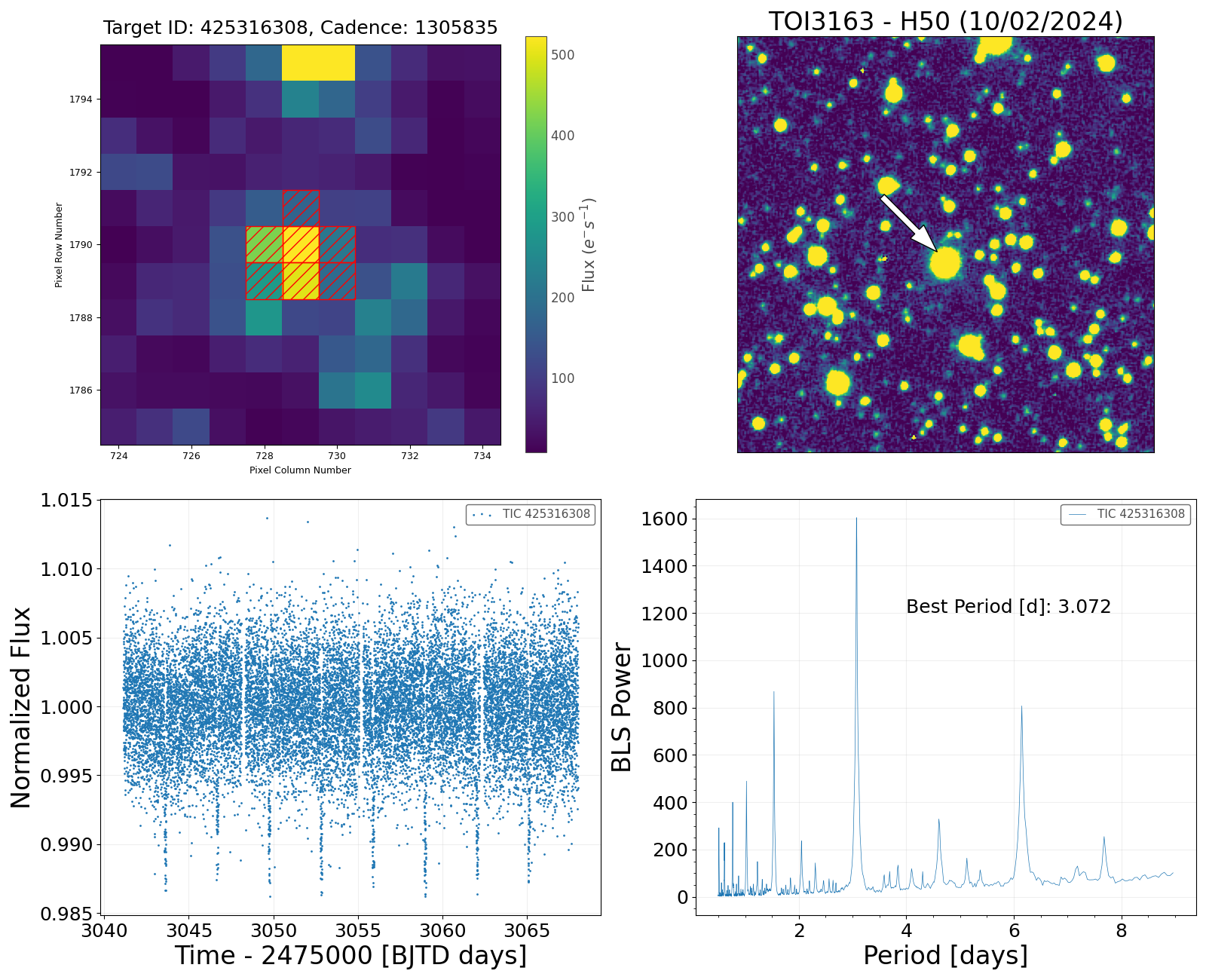}
        %\caption{}
        \label{fig:sptlcprd3163}
    \end{subfigure}
	\caption{Summary plots for observations of TOI3124 \textbf{(a)} and TOI3163 \textbf{(b)}. The TESS Sector 64 TPF, normalized light curves and periodogram plots are shown for each target. Cutouts of the stacked H50 images are also shown, matching the approximate orientation and scale of each TPF. In \textbf{(a)}, the orange light curve shows the raw, normalized flux for the star. The blue light curve is the detrended light curve (offset by a constant). The periodogram refers to the planetary period, based upon the detrended data.}
	\label{fig:tpflcprdmain}
\end{figure*}

TIC 425316308 is V $\approx$ 12.1 mag star in the constellation of Centaurus. The star was observed by TESS in Sectors 11 (23/04/19 -  20/05/19), 37 (02/04/21 - 28/04/21), 38 (29/04/21 - 26/05/21) and 64 (06/04/2023 - 04/05/2023). TOI3097 was originally alerted as a TOI on 04/06/2021 13:54:19 UTC by the Faint Star Search. Strong transit-like dips at roughly 1.5 \% are visible in the non-folded light curve. The period with the highest BLS power was 3.072 days, as can be seen in Figure \ref{fig:tpflcprdmain} (b). 
\newline\newline
TOI3163 was observed on 10/02/2024 with the H50 using the SDSS r' filter with exposures of 120 seconds. 90 images were obtained over an airmass range of 1.8 - 1.1. The sky was initially clear and moonless, but relatively strong winds lead to a poor median FWHM at 3.3". The baseline, ingress and midpoint were captured, but the egress was missed due to cloud. 

\subsubsection{TIC 442050698 (TOI4266)}
TIC 442050698 is a V $\approx$ 14.2 mag star in the constellation of Carina. The star was observed by TESS in Sectors 9 (28/02/19 - 25/03/19), 10 (25/03/19 - 22/04/19), 37 (02/04/21 - 28/04/21), 63 (10/03/23 - 06/04/23) and 64 (06/04/23 - 04/05/23). It was alerted as a TOI on 12/07/2021 14:48:48 UTC via the Faint Star Search. A possible transit signal is visible in the SPOC light curves, with an approximate period of 0.811 days.
\newline\newline
TOI4266 was observed by the H50 on 15/01/2024 with the SDSS r' filter with an exposure time of 150 seconds per image. The sky was clear and the moon illumination was low (at roughly 15 \%). Conditions were sub-optimal with very strong winds ranging from 15 - 20 km/h, resulting in a poor median FWHM of $3.6^{\prime\prime}$.  A total of 56 images were obtained over an airmass range of 1.1 - 1.3, covering baseline, ingress and egress.

\subsection{Spectroscopy}
To aid in the characterisation of the target stars and attempt to rule out double-lined binaries, we obtained mid-resolution reconnaissance spectra from the Wide Field Spectrograph \citep[WiFeS,][]{wifes} on the Australian National University (ANU) 2.3 m telescope. We used the R = 7000 red and blue gratings, which span the wavelength range of $\lambda$ = (420,700) nm. The velocity resolution of the instrument is $\delta$v = c/R $\approx$ 45 kms$^{-1}$. We took standard calibration data including bias, dome flats, twilight flats, Ne-Ar arc lamp and wire frames. Standards stars were also observed after each target for telluric corrections and flux calibration. Data were reduced using the PyWiFeS pipeline \citep{pywifes}. The observation logs for the spectroscopic measurements are shown in Table \ref{table:wifeslog}. 

\begin{table}[h!]
	\begin{center}
		\begin{tabular}{|p{1.0cm}|p{1.4cm}|p{1.2cm}|p{1cm}|p{1cm}|p{0.5cm}|}
			\toprule
			\headrow \textbf{ID} & \textbf{Dates [UTC]} & \textbf{Grating} & \textbf{Exp. Time [s]} & \textbf{Airmass} & \textbf{SNR} \\
			\midrule 
			\textbf{TOI3070} & 22/07/2024 & Red-7000
   \newline Blue-7000  & 120 & 1.85 & $\approx$ 5 \\
			\midrule 
			\textbf{TOI3097} & 23/07/2024 & Red-7000
   \newline Blue-7000 & 100 & 1.51 & $\approx$ 5\\
			\midrule
			\textbf{TOI3124} & 21/07/2024 & Red-7000
   \newline Blue-7000 & 50 & 2.01 & $\approx$ 5\\
			\midrule 
			\textbf{TOI3163} & 10/07/2024 & Red-7000
   \newline Blue-7000 & 50 & 1.12 & $\approx$ 5\\
			\bottomrule 
		\end{tabular}
		\caption{Observation logs for TASSIE targets using the WiFeS spectrograph on the ANU 2.3 m telescope at Siding Spring Observatory.}
		\label{table:wifeslog}
	\end{center}
\end{table}

\section{Data Analysis and Results}
\subsection{Stellar Parameters}
\subsubsection{Spectral Energy Distribution Fitting}
\begin{figure}[hb!]
	\centering
	\begin{subfigure}
		\centering
		\includegraphics[width=0.95\linewidth, height = 0.65\linewidth]{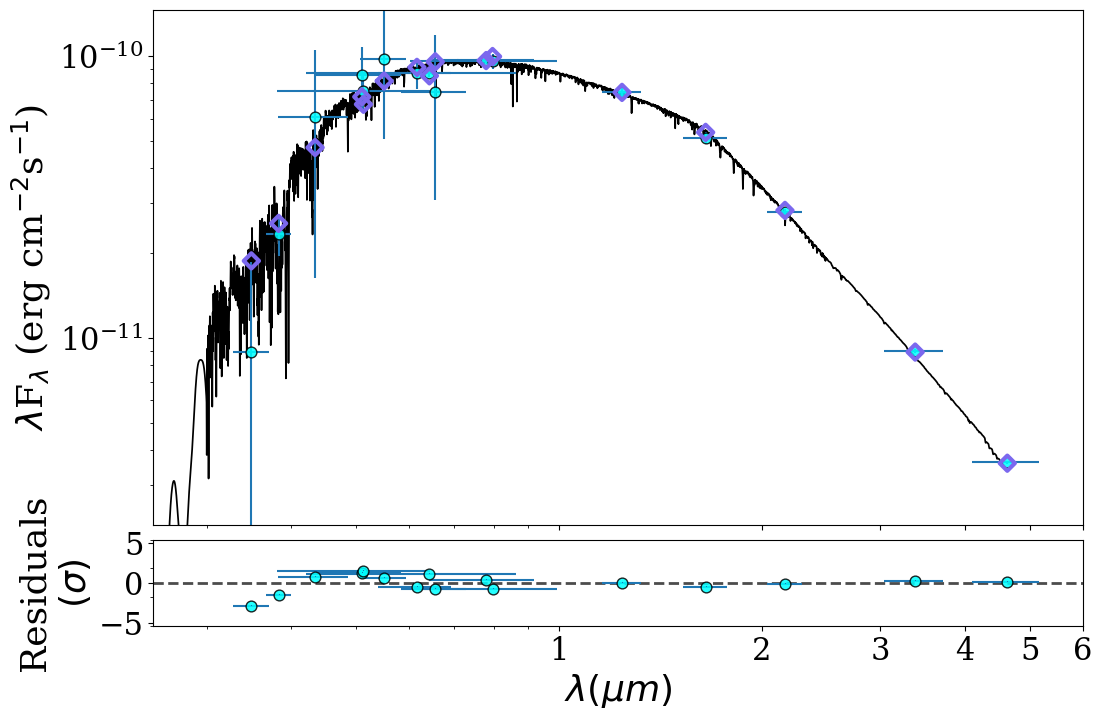}
	\end{subfigure}
	~
	\begin{subfigure}
		\centering
		\includegraphics[width=0.95\linewidth, height = 0.9\linewidth]{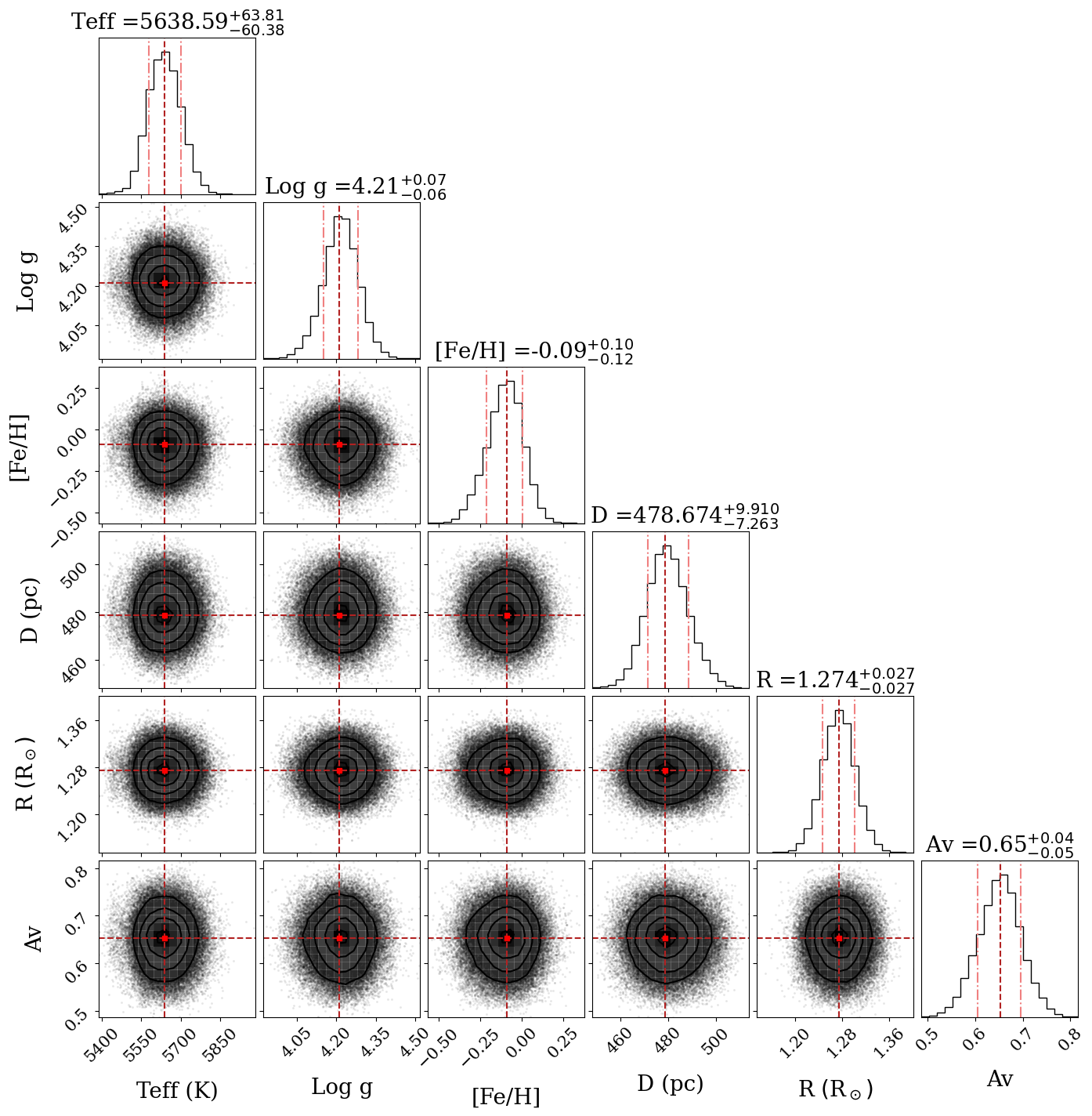}
	\end{subfigure}
	\caption{The results of SED fitting for TOI3124 with ARIADNE. \textbf{(a)} The spectral energy distribution of TOI3124 with Phoenix V2 model overlaid. \textbf{(b)} The corner plot for stellar parameters of TOI3124.}
	\label{fig:SED3124}
\end{figure}

As a first step in determining the stellar parameters, we used the spectral energy distribution (SED) fitting code ARIADNE \citep{ariadne}. Four model grids were utilised in our analysis, including Phoenix v2 \citep{PhoenixV2}, Kurucz 1993, ATLAS9 \citep{CastelliKurucz} and BT-Settl \citep{BTmodels}. We queried Vizier for all publicly available photometry from various surveys including \gaia DR3, 2MASS \citep{2MASS} and SkyMapper \citep{skymapper}. We set broad normal priors on the effective temperature ($T_{eff}$), surface gravity ($\log{(g)}$), metallicity ([Fe/H]) and extinction ($A_{v}$, corrected from $A_{0}$) centred on the reported \gaia DR3 values from XP spectra \citep{dr3}. As the uncertainties are often underestimated, we used empirically measured precisions of $\sigma_{log(g)}$ = 0.1 dex and $\sigma_{[Fe/H]}$ = 0.2 dex from the DR3 documentation\footnote{\url{https://gea.esac.esa.int/archive/documentation/GDR3/Data_analysis/chap_cu8par/sec_cu8par_validation/ssec_cu8par_qa_atmospheric-aps.html}} and set 10 $\%$ errors on $T_{eff}$ and $A_{v}$ as prior widths. The \gaia DR3 parallaxes with uncertainties were provided to constrain the distances.
\newline\newline
The radius and distance of each star is determined by scaling the stellar model grid surface fluxes to the observed fluxes \citep[see Equation 1 of][]{ariadne}. The final stellar parameters come from Bayesian model averaging of individual model posteriors. MESA Isochrones and Stellar Tracks \citep[MIST, ][]{mist} models were then used to estimate the mass by interpolation. We ran nested sampling with 1000 live points and 100000 samples, with evidence tolerance of $\Delta\log(Z)$ = 0.05. An example of the results of this fitting process are shown in Figure \ref{fig:SED3124}. It is worth noting some ambiguity in the results due to uncertain distances and possible multiplicity for TOI3070 and TOI4266, as discussed in Section 4.2.

\subsubsection{Semi-Empirical Relations}
We excluded TOI3097 and TOI3163 from the SED fitting, due to strong contamination of ground-based photometry. For TOI3163, there is a moderately bright (G $\approx$ 15) non-companion source at roughly 3" away. TOI3097 has already been noted to have a bound wide companion. We instead estimate stellar parameters using the semi-empirical relations of \citet{Torres2010} for early-to-mid type dwarfs and \citet{Mann2015} for late dwarfs in the range of K7-M7. For TOI3097A and TOI3163, we adopt estimates for effective temperature, surface gravity and metallicity from \gaia DR3 \citep{dr3} as listed in Table \ref{table:stellparams}. We find mass and radii of (0.63 $\pm$ 0.04 $M_{\odot}$, 0.87 $\pm$ 0.03 $R_{\odot}$) and (1.29 $\pm$ 0.08 $M_{\odot}$, 1.56 $\pm$ 0.05 $R_{\odot}$) for TOI3097A and TOI3163, respectively. 
\newline\newline 
For TOI3097B, we used the BPRP-$T_{eff}$ relation of Mann et al. (2015) to estimate the effective temperature of $T_{eff}$ = 3940 $\pm$ 223 K. We then used the radius-$T_{eff}$ relation to estimate a radius of 0.30 $\pm$ 0.02 $R_{\odot}$. We note that \gaia reports a low metallicity of [Fe/H] $\approx$ -1.0 for the TOI3097 system, implying these are subdwarf stars. As the \citet{Mann2015} sample does not extend below [Fe/H] = -0.6 dex, we also checked the radii using data from \citet{kesseli2019}. We refit a colour-radii relation for subdwarfs (as discussed in the Appendix) and find a matching radius of 0.28 $R_{\odot}$. As subdwarf masses have been scarcely measured, we can only roughly estimate mass and surface gravity using the dwarf sequence of \citet{Mamajek2013}. Interpolating with radius, we find M = 0.27 $\pm$ 0.03 $M_{\odot}$ and log(g) = 4.91 $\pm$ 0.08.

\subsubsection{Template Matching}
As an independent check of the properties of the TOIs, we now investigate the WiFeS spectra. To assign stellar classifications, we utilise PyHammer v2 \citep[see][]{pyhammerv1, pyhammerv2}. By comparing multiple spectral indices and colour regions measured from Sloan Digital Sky Survey 
\citep{sdss} spectral templates with those measured from the input spectra, spectral types and metallicity are estimated through the minimization of a $\chi^2$ difference. An advantage of this method is robustness against flux calibration errors and low SNR spectra. The code can also identify some double-lined binaries. Due to the low radial velocity precision and limited wavelength coverage, we shifted the WiFeS spectra to the rest wavelength by assuming the \gaia DR3 radial velocities. The estimated values were consistent with those found from the previous methods. However, the low SNR of data prevented further refinement of the stellar parameters.

\subsubsection{Kinematics}
The kinematics of stars can provide information upon the system age and possible associations of the star within the galaxy \citep{bensby2003}. We calculate the galactic velocities of the five TOIs using the equations defined by \citet{soderblom1987}, with updated transformation matrix for the J2000 epoch. We adopt the parallax, proper motions and radial velocities in Table \ref{table:stellparams} from \gaia DR3 \citep{dr3}. We take the solar velocities of ($U_{\odot}, V_{\odot}, W_{\odot}$) = (8.5,13.38,6.49) km/s from \citet{cos2011}. Here, U is defined as positive towards the centre of the galaxy, V is positive with the direction of galactic rotation and W is positive in the direction of the North galactic pole. The results of this analysis are listed in Table \ref{table:stellparams}.
\newline\newline
Using the membership probabilities defined by \citet{bensby2003}, we place four of the TOIs in the thin disk. However, the kinematics of TOI3097 (specifically the large negative value of W and positive value of V) potentially indicate a thick disk star, with a relative thick disk to thin disk probability of $\approx$ 1.6. This is consistent with the low metallicity measured. We also utilise the BANYAN $\Sigma$ tool \citep{banyan} to check if any of the targets are members of known young stellar associations. We obtain a 99.9 \% probability of the targets being field stars.

\subsubsection{Stellar Activity}
\begin{figure}[h!]
	\centering
\includegraphics[width=0.99\linewidth, height = 0.99\linewidth]{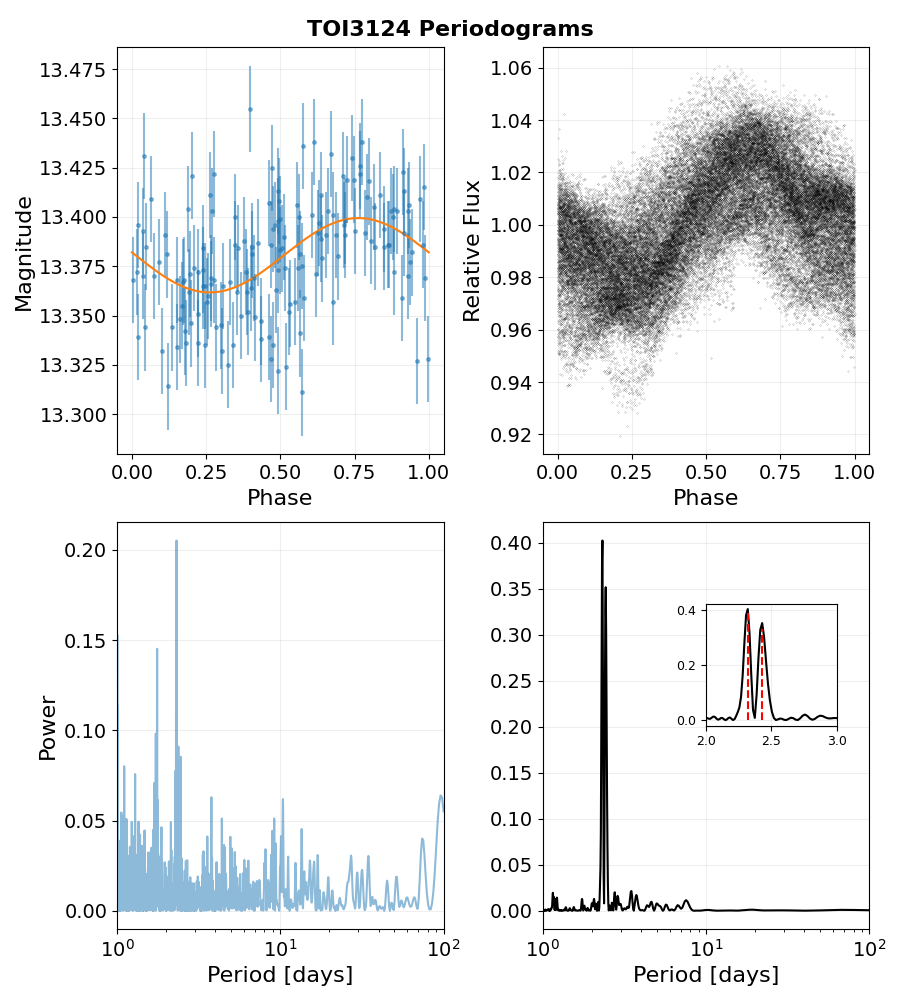}
	\caption{The ASAS-SN (in blue) and TESS (in black) folded light curves and periodogram for TOI3124. The inset shows the peak power periods, closely spaced at 2.31 and 2.43 days.}
	\label{fig:tessperiodogram}
\end{figure}

Stellar activity can add significant complexity to analysis of transit signals. However, it can be used to infer the stellar rotation period which is often used as a proxy for age \citep[e.g.,][]{mamajek2008}. As TESS observes in cycles of roughly 27 days per sector, rotational periods longer than this cannot be not well-measured. We therefore also investigate public archival photometry from the ASAS-SN \citep{asas-sn17} and ATLAS \citep{ATLAS_2018} surveys. We search for periodicity using a Lomb-Scargle periodogram search, calculating the False Alarm Probability (FAP) for the maximum power period. Aside from TOI3124, we do not find any consistent periods in the three data sets of significance. 
\newline\newline
It is apparent from the TESS and ASAS-SN light curves that TOI3124 is an active star at the 5-10 \% level. We attribute at least part of this variability to stellar spots. This conclusion comes from the quasi-periodic and morphologically variable modulation, along with the low effective temperature. The latter rules out the possibility of most pulsators, such as $\delta$ Scuti and $\gamma$ Doradus, which are usually F-type or earlier \citep{reinhold2013, balona2016}. We also find evidence of at least one flaring event in the Sector 64 light curve (at BJTD $\approx$ 3044.38), further indicating this is an active star. Inspecting the periodogram from each survey in Figure \ref{fig:tessperiodogram}, we find a matching period of 2.31 $\pm$ 0.01 days. A marginally weaker signal is seen in the TESS periodogram at 2.43 days. This is unusually quick for a mid G-dwarf and may be related to multiplicity in the system.

\begin{table*}[h!]
	\begin{center}
		\begin{tabular}{|p{2.5cm}|p{2.5cm}|p{2.5cm}|p{2.5cm}|p{2.5cm}|p{2.5cm}|} 
			\toprule
			\headrow \textbf{Parameters} & \textbf{TOI3070}&\textbf{TOI3097} & \textbf{TOI3124} & \textbf{TOI3163} & \textbf{TOI4266}  \\
			\midrule 
			\textbf{Catolog IDs} & TIC 452049244 \newline 2MASS \newline J11333579-5642086 \newline \gaia-DR3 5342871047750973312 & TIC 163262555 \newline 2MASS \newline J11321797-4729477 \newline \gaia-DR3 5375076739733884544 & TIC 454918388 \newline 2MASS \newline  J11523546-7725020 \newline \gaia-DR3 5224239068846508544 & TIC 425316308 \newline 2MASS \newline J12442132-5500479 \newline \gaia-DR3 6073589158249558400 &
			TIC 442050698 \newline 2MASS \newline J09430690-5830389 \newline \gaia-DR3 5305817570291360128\\
			\hline\hline 
			\headrow \multicolumn{6}{|c|}{\textbf{Measured Properties}}\\
			\hline\hline 
            \textbf{\gaia G} [mag] & $13.78^{+0.01}_{-0.01}$ & A: $13.78^{+0.01}_{-0.01}$ \newline B: $15.19^{+0.01}_{-0.01}$ & $13.60^{+0.01}_{-0.01}$ & $12.21^{+0.01}_{-0.01}$ & $14.01^{+0.01}_{-0.01}$\\
            \hline
			\textbf{V} [mag]  & $14.28^{+0.10}_{-0.10}$ & $13.80^{+0.02}_{-0.02}$ (*) & $13.28^{+0.05}_{-0.05}$ & $12.05^{+0.14}_{-0.14}$ (*) & $14.21^{+0.01}_{-0.01}$ \\
	        \hline 
			\textbf{Ks} [mag]  & $12.39^{+0.03}_{-0.03}$ & $11.56^{+0.03}_{-0.03}$ (*) & $11.30^{+0.023}_{-0.023}$ & $10.96^{+0.03}_{-0.03}$ (*) & $12.07^{+0.02}_{-0.02}$ \\
			\hline
            \textbf{\gaia BP-RP} [mag]  & $0.80^{+0.01}_{-0.01}$ & A: $1.14^{+0.01}_{-0.01}$ \newline B: $1.81^{+0.2}_{-0.2}$ & $1.05^{+0.01}_{-0.01}$ & $0.75^{+0.01}_{-0.01}$ & $1.06^{+0.01}_{-0.01}$ \\
			\hline
			\textbf{Parallax} [mas] & $0.465^{+0.042}_{-0.042}$ & $2.408^{+0.018}_{-0.018}$ & $2.036^{+0.012}_{-0.012}$ & $1.805^{+0.012}_{-0.012}$ & $1.125^{+0.052}_{-0.052}$ \\ 
			\hline
			\textbf{PM}$_{RA}$ \newline [mas $yr^{-1}$] & $-1.320^{+0.043}_{-0.043}$ & $-16.449^{+0.013}_{-0.013}$ & $-29.769^{+0.013}_{-0.013}$ & $-13.073^{+0.008}_{-0.008}$ &$-9.678^{+0.063}_{-0.063}$ \\
			\hline
			\textbf{PM}$_{DEC}$ \newline [mas $yr^{-1}$] & $-0.527^{+0.039}_{-0.039}$ & $25.469^{+0.014}_{-0.014}$ & $2.477^{+0.014}_{-0.014}$ & $-1.893^{+0.011}_{-0.011} $& $8.311^{+0.066}_{-0.066}$ \\
			\hline\hline
		    \headrow \multicolumn{6}{|c|}{\textbf{Derived Properties}}\\
		    \hline\hline 
			\textbf{Extinction} $A_{V}$ [mag] & $0.47^{+0.04}_{-0.04}$ (\textdagger) & $0.01^{+0.01}_{-0.01}$ & $0.65^{+0.07}_{-0.08}$ (\textdagger) & $0.38^{+0.04}_{-0.04}$ & $0.72^{+0.08}_{-0.08}$ (\textdagger) \\
			\hline
			\textbf{Effective Temp.} [K] & $6293^{+104}_{-104}$ (\textdagger) & A: $4802^{+100}_{-100}$ \newline\newline  B: $3940^{+223}_{-223}$ & $5639^{+104}_{-100}$ (\textdagger) & $6501^{+100}_{-100}$ &  $5637^{+111}_{-98}$ (\textdagger) \\
			\hline
			\textbf{log($\frac{g}{[cms^{-2}]}$)}  & $4.09^{+0.08}_{-0.09}$ (\textdagger) & A: $4.36^{+0.10}_{-0.10}$ \newline\newline B: $4.91^{+0.10}_{-0.10}$ & $4.21^{+0.11}_{-0.10}$ (\textdagger) & $4.16^{+0.10}_{-0.10}$ &  $4.33^{+0.09}_{-0.09}$ (\textdagger) \\
			\hline
			\textbf{Radius} [$R_{\odot}$] & $1.79^{+0.09}_{-0.08}$ (\textdagger) & A: $0.87^{+0.03}_{-0.03}$ \newline\newline B: $0.30^{+0.02}_{+0.02}$ & $1.27^{+0.05}_{-0.05}$ (\textdagger) & $1.56^{+0.05}_{-0.05}$  &  $1.15^{+0.04}_{-0.03}$ (\textdagger) \\
			\hline
			\textbf{Mass} [$M_{\odot}$] & $1.33^{+0.08}_{-0.08}$ (\textdagger) & A: $0.63^{+0.04}_{-0.04}$ \newline\newline B: $0.27^{+0.08}_{-0.08}$ &  $0.93^{+0.11}_{-0.06}$ & $1.29^{+0.08}_{-0.08}$ & $0.87^{+0.06}_{-0.03}$ (\textdagger) \\
			\hline
			\textbf{Distance} [pc] & $1215^{+50}_{-48}$ (\textdagger) & $415^{+3}_{-3}$ & $479^{+15}_{-13}$ (\textdagger) & $554^{+4}_{-4}$ & $617^{+16}_{-12}$ (\textdagger) \\
			\hline
			\textbf{[Fe/H]} (dex) & $-0.05^{+0.10}_{-0.08}$ (\textdagger) & $-1.07^{+0.20}_{-0.20}$ & $-0.09^{+0.16}_{-0.19}$ (\textdagger) & $-0.22^{+0.20}_{-0.20}$ & $-0.37^{+0.09}_{-0.08}$ (\textdagger) \\
			\hline
			\textbf{Spectral Type} & F5 - F7 V & A: K3 V(I) \newline B: M0 - M1 V(I) & G4 - G6 V & F4-F6 V & G4 - G6 V \\
            \hline 
            \textbf{Rot. Period} [days] & - & - & 2.32 $\pm$ 0.01 & - & - \\
			\hline\hline
			\headrow \multicolumn{6}{|c|}{\textbf{Kinematic Properties}}\\
			\hline\hline 
			\textbf{RV} [$kms^{-1}$] & $45.35^{+4.70}_{-4.70}$ & $-31.38^{+2.79}_{-2.79}$ & $12.36^{+5.25}_{-5.25}$ & $-7.25^{+2.88}_{-2.88}$ & $-6.88^{+4.93}_{-4.93}$ \\
			\hline
			\textbf{U} [$kms^{-1}$] & $14.6^{+2.0}_{-2.0}$ & $-11.6^{+0.9}_{-0.9}$ & $-47.1^{+2.5}_{-2.5}$ & $-23.9^{+1.5}_{-1.5}$ & $-48.0^{+1.8}_{-1.8}$ \\
			\hline 
			\textbf{V} [$kms^{-1}$] & $-34.0^{+4.0}_{-4.0}$ & $24.6^{+2.6}_{-2.6}$ & $-29.0^{+4.0}_{-4.0}$ & $0.7^{+2.4}_{-2.4}$ & $10.0^{+5.0}_{-5.0}$ \\
			\hline
			\textbf{W} [$kms^{-1}$] & $0.1^{+0.8}_{-0.8}$ & $-56.9^{+0.8}_{-0.8}$ & $-7.0^{+1.4}_{-1.4}$ & $0.4^{+0.4}_{-0.4}$ & $6.8^{+0.5}_{-0.5}$ \\
            \bottomrule 
		\end{tabular}
		\caption{Stellar parameters for TASSIE TOI host stars. Measurements with (*) indicate suspected blended photometry. Derived properties with (\textdagger) are from SED-fitting, with the remainder combining \gaia DR3 and semi-emperical relations.}
		\label{table:stellparams}
	\end{center}
\end{table*}

\subsection{Planet Validation}
In order to validate these planet candidates, we will need to rule out other astrophysical scenarios that could explain the transit-like features. As we have detected these signals with the H50, we can dismiss the possibility of TESS systematics creating the signal. We can also exclude many potential host stars, due to our much higher resolution of 0.8 $^{\prime\prime}$/pixel. We now explore other methods to exclude false positive scenarios.

\subsubsection{Astrometry}
Eclipsing binaries can mimic the signatures of transiting planets. 
One method for identifying the presence of bound companions that could introduce the transit-like signal is through checking the quality of the \gaia astrometric solution. To do this we look at the `Renormalized Unit Weight Error' (RUWE), which is defined as:
\begin{equation}
    RUWE = \frac{1}{f} \sqrt{\frac{\chi^2}{(n - m)}}
\end{equation}
here, n is the number of good observations, m is the number of fit parameters and f is a reference value which is a function of magnitude and colour. Generally, sources with RUWE > 1.4 are considered to have questionable astrometry, potentially indicating the presence of multiplicity in the system \citep{ruwe}. TOI3070 and TOI4266 are notable as they have RUWE values of 3.31 and 5.03 respectively, indicating very poor astrometric fits. TOI3163 is also interesting as the brightest star has a RUWE = 0.8, but a close background star has a value of 2.8.  We must therefore be cautious with the quoted values (such as parallax and proper motions) for these stars. The other stars have RUWE values below the threshold.

\subsubsection{Archival Imaging}
To help rule out background EBs, we retrieve archival images from the 2MASS survey \citep{2MASS}. These range from 20 to 25 years prior to imaging with the H50. We extract catalogues from our stacked images using SExtractor \citep{sex} and plot the detected sources over the 2MASS images. Due to the generally low proper motions, the positions of the targets do not change drastically across the two epochs. No contaminating stars were identified which were not already flagged from our own imaging.

\subsubsection{Statistical Testing with TRICERATOPS}
We perform statistical tests for false positive scenarios using TRICERATOPS \citep{Triceratops}. We utilise the TESS light curves and transit depth estimates, along with the stellar parameters from Section 4.1 in this analysis. We remove many surrounding stars from the calculations due to the transit-like signals detected with the H50, further tightening the constraints. The output of this process are relative probabilities of various scenarios, along with estimates of the false positive probability (FPP) and nearby false positive probability (NFPP). We repeated the fitting process 10 times on each TOI to ensure the FPPs and NFPPs were consistent. We list these estimated values and uncertainties in Table \ref{table:triresults} for each candidate. 
\newline\newline
We can see that TOI3070 has a very high FPP of $\approx$ 0.79, with the most likely scenario being an EB (either around the primary or a secondary star). The high RUWE value also supports this hypothesis. Interestingly, the SED fitting process produced odd radius values for the estimated spectral type (F5 - F7 V), ranging from 1.8 to 3 $R_{\odot}$ instead of the expected 1.3 - 1.4 $R_{\odot}$ \citep{Mamajek2013}. Similarly for TOI4266, we have a significant FPP of $\approx$ 0.84, with the most probable scenario being an EB (nearby or with double period). Once again, this is consistent with the high RUWE value. We shall exclude these candidates from further modelling, as we are interested in planets only. Finally, we note that whilst the mean FPP for TOI3124 is very high at $\approx 0.99$ (making the signal a likely false positive), we wanted to check that this result was unbiased by the method of activity detrending. Hence, we decided to include it in the modelling stage. 

\begin{table}[h!]
	\begin{center}
		\begin{tabular}{|p{1cm}|p{2cm}|p{2cm}|p{1.8cm}|} 
			\toprule
			\headrow \textbf{ID} & \textbf{FPP} & \textbf{NFPP} & \textbf{Most Likely Scenario} \\
			\midrule
			TOI3070 & 0.786 $\pm$ 0.007 & < 0.001 & EB \\ 
            TOI3097 & 0.060 $\pm$ 0.002 & 0.003 $\pm$ 0.001 & TP \\ 
            TOI3124 & 0.989 $\pm$ 0.003 & < 0.001 & SEB \\ 
            TOI3163 & 0.229 $\pm$ 0.118 & < 0.001 & TP \\ 
            TOI4266 & 0.843 $\pm$ 0.006 & 0.5846 $\pm$ 0.0073 & NEB or EB (2xP) \\ 
			\bottomrule
		\end{tabular}
\caption{TRICERATOPS Results for TASSIE TOIs}
\label{table:triresults}
\end{center}
\end{table}

\subsection{Transit modelling}
We model the transit light curves using the Juliet package \citep{Juliet}, which combines nested sampling algorithms with the light curve calculations of BATMAN \citep{batman}. For each of the remaining candidates, we perform joint modelling with high cadence (i.e exposures of 200 s or less) TESS data and light curves from the H50. We choose to incorporate Gaussian Processes (GPs) into our modelling to account for any stellar activity or remaining systematics. To estimate the quadratic limb darkening coefficients for each star, we use the ExoCTK limb darkening calculator\footnote{\url{https://exoctk.stsci.edu/limb_darkening}} with Phoenix V2 \citep{PhoenixV2} models. We then convert to the limb darkening parameterization suggested by \citet{kipping2013}, and set a normal prior with a conservative error of 10 \%. We use the flux ratio calculated by TRICERATOPS to estimate the dilution factor for the H50 light curves. The SPOC light curves are already corrected for contamination, so we set the dilution factor to 1 for TESS data.  We list descriptions of parameters and priors for the modelling in the Appendix in Table \ref{table:priors} \citep[for further details, see][]{espinoza2018}.

\subsubsection{TOI3097}
Due to this being a binary system, we started with a coarse search of the parameter space with no constraint on the stellar density. As we can use the light curve to recover the stellar density from Kepler's law, we use this to check which star the planet candidate is orbiting. It was found that the brighter, primary star is the likely host. We then remodelled the light curves, this time using the stellar density constraint of $\rho = 1347.1 \pm 490.6$ kg/$m^3$ (3$\sigma$ error) and incorporating an (approximate) Matern kernel GP of the form:
\begin{equation}
k(\tau) = \sigma_{GP}^{2} [(1+\frac{1}{\epsilon}) \exp^{-(1-\epsilon) s_{i}} + (1-\frac{1}{\epsilon}) \exp^{-(1+\epsilon) s_{i}}]
\end{equation}
where, $s_{i}$ = $3\tau/\rho_{GP}$ is a scaled timescale of the GP and $\sigma_{GP}$ relates to the amplitude of the GP. The variable $\epsilon$ is set to 0.01 (see \citet{celerite} and \citet{Juliet} for more discussion of the interpretation and use of this kernel).
We ran nested sampling with 1000 live points, initially assuming a circular orbit. We also performed a separate run adding eccentricity (e) and argument of periastron ($\omega$), using the parameterization of $\sqrt{e} sin(\omega)$ and $\sqrt{e} cos(\omega)$. However, we found this produced a worse fit with a lower log-likelihood ($\Delta lnZ = -2.276$). The measured and derived planetary parameters are shown in Table \ref{table:planetresults}, being consistent with a sub-Jovian to Jovian size planet on a circular orbit.

\begin{figure}[h!]
    \centering
    \includegraphics[width=0.99\linewidth, height = 1.2\linewidth]{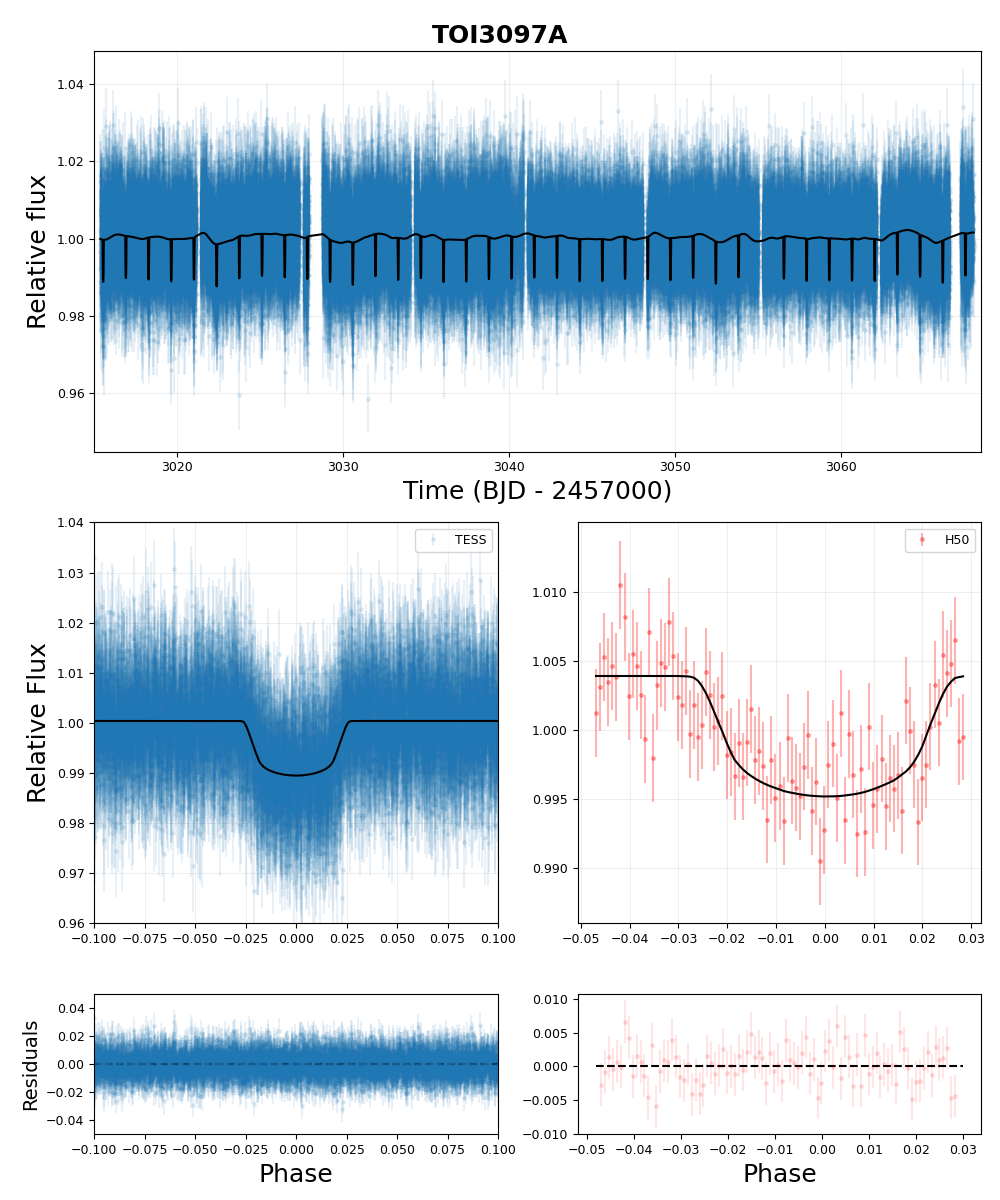}
    \caption{The light curve and joint transit model for TOI3097Ab. The top plot shows the full Sector 63 \& 64 data from TESS, with the best-fit transit + GP model overlaid. In the bottom left plot, we show the phase-folded TESS data, model and residuals. On the bottom right plot, we show data and residuals from the H50 telescope.}
    \label{fig:toi3097Ab}
\end{figure}

\subsubsection{TOI3124}
The stellar variability required experimentation with the form of the GP used for this target. Like TOI3097, we started with a default Matern 3/2 kernel. We also tried a quasi-periodic kernel introduced by \citet{celerite}, with four hyperparameters relating to the amplitude, shifting, length-scale and period of the GP. The advantage of this kernel is it is physically motivated by the quasi-periodic signals seen in stellar activity. We ran nested sampling with 1000 live points, again assuming a circular orbit and then relaxing this in a latter modelling run. Both runs resulted in unphysical values for the impact parameter (b > 1) and radius ratio (p > 0.2), with strong degeneracy between the two. Combining this with the proposed `planetary' period matching the activity period of the star, we conclude that this signal is non-planetary in nature. We either attribute this to small-scale, long-lived star spots or a grazing eclipsing binary. This interpretation is supported by the absence of a clear transit signal in more recent V band observations from the H50 (shown in the Appendix). Roughly achromatic transit depths are expected for true planetary signals.

\subsubsection{TOI3163}
\begin{figure}[h!]
    \centering
    \includegraphics[width=0.99\linewidth, height = 1.2\linewidth]{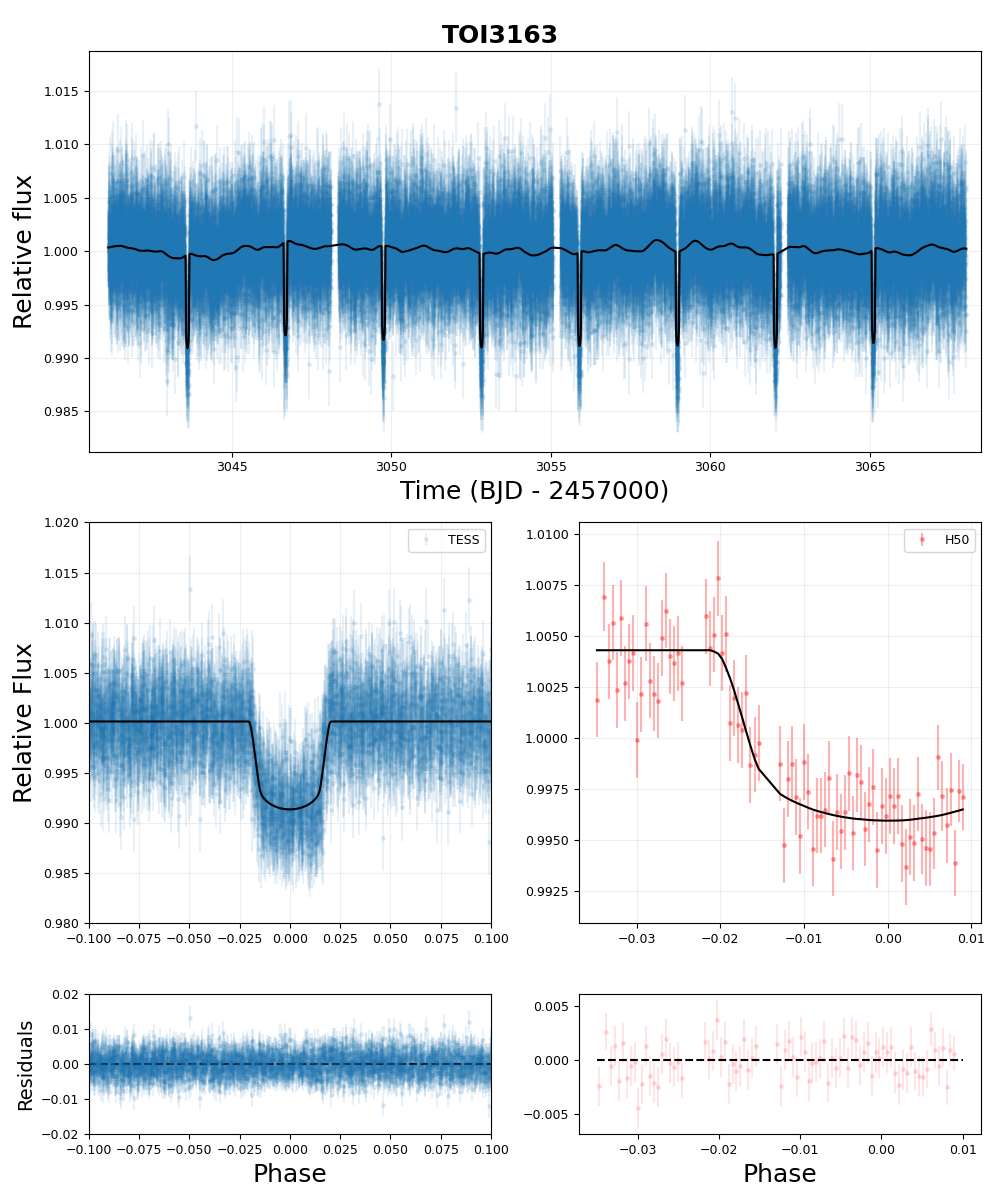}
    \caption{The light curve and joint transit model for TOI3163b. The top plot shows the full Sector 64 data from TESS, with the best-fit transit + GP model overlaid. In the bottom left plot, we show the phase-folded TESS data, model and residuals. On the bottom right plot, we show data and residuals from the H50 telescope.}
    \label{fig:toi3163b}
\end{figure}

Finally, we modelled the data for TOI3163 using a prior on the stellar density of $\rho$ = 478.4 $\pm$ 164.2 kg/$m^3$ (3$\sigma$ error). The light curve is shown in Figure \ref{fig:toi3163b}, with the results listed again in Table \ref{table:planetresults}. The posterior distributions are shown in the Appendix. We find a Jovian size planet with $R_{3163b} = 1.42 \pm 0.05 \, R_{J}$ and $P_{3163b} = 3.074966 \pm 0.000022$ days. A circular orbit model is preferred, with a log-likelihood difference for the eccentric model of $\Delta lnZ = -7.373$. 

\begin{table}[h!]
	\begin{center}
		\begin{tabular}{|p{2.5cm}|p{2.5cm}|p{2.5cm}|} 
			\toprule
			\headrow \textbf{Parameters} & \textbf{TOI3097Ab} & \textbf{TOI3163b}\\
            \midrule
			\headrow \multicolumn{3}{|c|}{\textbf{Fitted Parameters}} \\
            \midrule
			$P_{b}$ [days] & 1.368386 $\pm$ 0.000006 & 3.074966 $\pm$ 0.000022\\ 
            $T_{0}$ [BJD-2457000] & 3015.5360 $\pm$ 0.0005 & 3043.6007 $\pm$ 0.0006\\ 
            $R_{p}/R_{*}$ & 0.103 $\pm$ 0.003 & 0.0915 $\pm$ 0.0013 \\ 
            $b$ & $0.71^{+0.06}_{-0.09}$ & $0.66^{+0.05}_{-0.05}$\\ 
            $a/R_{*}$ & 5.2 $\pm$ 0.5 & 6.9 $\pm$ 0.4\\
            $\rho_{*}$ [kg/$m^{3}$] & $1377^{+460}_{-347}$ & $657^{+107}_{-103}$\\
            e & 0 (fixed) & 0 (fixed) \\
            $\omega$ [deg] & 90 (fixed) & 90 (fixed) \\
            \midrule
			\headrow \multicolumn{3}{|c|}{\textbf{Derived Properties}} \\
            \midrule
            $R_{p}$ [$R_{J}$] & 0.89 $\pm$ 0.04 & 1.42 $\pm$ 0.05 \\
            i [deg] & $82.1 \pm 1.6$ & $84.5 \pm 0.7$\\
            a [au] & 0.021 $\pm$ 0.002 & $0.050 \pm 0.003$ \\
            $T_{eq}$ ($A_{B}$ = 0) [K] & $1960 \pm 110$ & $3090 \pm 110$\\
            F [$W m^{-2}$] & $(1.1 \pm 0.3) \, \times \, 10^{6}$ & $(2.1 \pm 0.3) \, \times \, 10^{6}$  \\
            \midrule 
			\headrow \multicolumn{3}{|c|}{\textbf{Instrumental Parameters}} \\
            \midrule 
            $M_{TESS}$ ($10^{-5}$) & $-38 \pm 12$ & $-1.0 \pm 0.7$ \\
            $J_{TESS}$ [ppm] & $3.69^{+42.15}_{-3.37}$ & $3.55^{+30.94}_{-3.22}$ \\
            $q_{1, TESS}$ & 0.34 $\pm$ 0.03 & 0.22 $\pm$ 0.02\\
            $q_{2, TESS}$ & 0.40 $\pm$ 0.04 & 0.35 $\pm$ 0.04\\
            $D_{H50}$ & 0.80 $\pm$ 0.01 & $0.94 \pm 0.01$ \\
            $M_{H50}$ & $-0.005^{+0.038}_{-0.004}$ & $-0.004^{+0.022}_{-0.002}$ \\
            $J_{H50}$ [ppm] & $6.13^{+120.22}_{-5.72}$ & $5.54^{+101.01}_{-5.20}$\\
            $q_{1, H50}$ & 0.47 $\pm$ 0.05 & 0.32 $\pm$ 0.03\\
            $q_{2, H50}$ & 0.43 $\pm$ 0.04 & 0.38 $\pm$ 0.04 \\
            \midrule
			\headrow \multicolumn{3}{|c|}{\textbf{GP Parameters}} \\
            \midrule
            $\sigma_{GP, TESS}$ ($10^{-4}$) & $8.0^{+0.9}_{-0.8}$ & $4.1^{+0.5}_{-0.4}$ \\
            $\rho_{GP, TESS}$ & $0.509^{+0.132}_{-0.102}$ & $0.296^{+0.085}_{-0.067}$\\ 
            $\sigma_{GP, H50}$ & $0.005^{+0.052}_{-0.005}$ & $0.004^{+0.040}_{-0.003}$ \\ 
            $\rho_{GP, H50}$ & $11.602^{+224.535}_{-11.338}$ & $0.265^{+6.275}_{-0.224}$ \\ 
			\bottomrule 
		\end{tabular}
		\caption{The modelling results from Juliet for the two planets.}
		\label{table:planetresults}
	\end{center}
\end{table}

\section{Discussion}
\subsection{The Short-Period Planet Population}
\begin{figure}[hb!]
    \centering
    \includegraphics[width=0.99\linewidth]{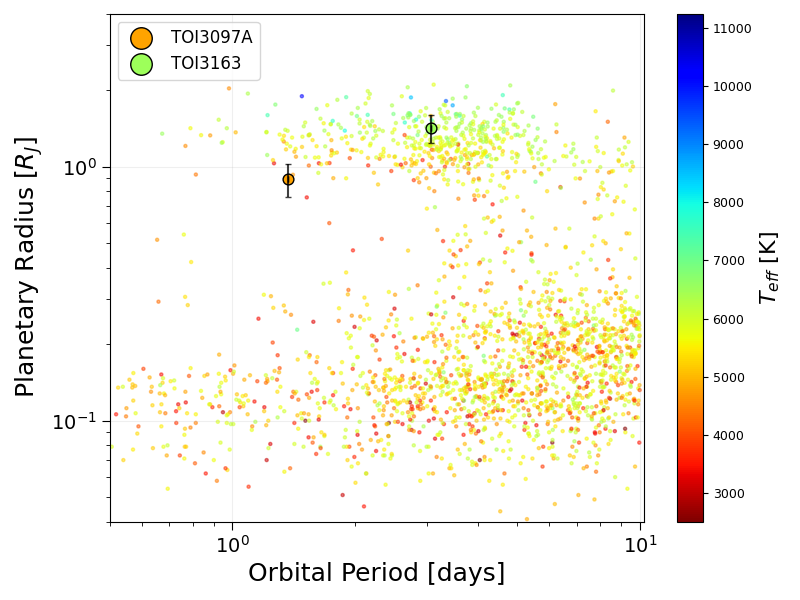}
    \caption{The radius-period plane for short period exoplanets, coloured by stellar effective temperature. TOI3097Ab and TOI3163b are shown as points with error bars (corresponding to 3$\sigma$).}
    \label{fig:shortperiod}
\end{figure}

To put these two candidates into context, we show the short-period exoplanet population in Figure \ref{fig:shortperiod}. We see that TOI3163 is likely a classical Hot Jupiter, as it falls in the middle of the distribution of similar planets. More interestingly, TOI3097Ab leans towards being a Saturn-size planet. Inspecting the position of TOI3097Ab in the radius-period plane, we note that it is on the upper edge of the sub-Jovian desert. This boundary is expected to be sculpted by photoevaporation, which will be testable through the detection of atmospheric loss from desert planets with the upcoming Ariel mission \citep{ariel, savanna}. TOI3097Ab could also conform to the high-eccentricity tidal migration scenario with the presence of the wide M sub-dwarf companion acting to perturb the planet into an eccentric orbit. The current low eccentricity would then be consistent with tidal circularization \citep{DawsonJohnson2018}. Combining this information with the potentially low metallicity suggests that the TOI3097 system will be an interesting target for further characterisation with mass measurements and transmission spectroscopy.

\subsection{Search for Transit Timing Variations}
\begin{figure}[ht!]
	\centering
	\begin{subfigure}
		\centering
		\includegraphics[width=0.96\linewidth]{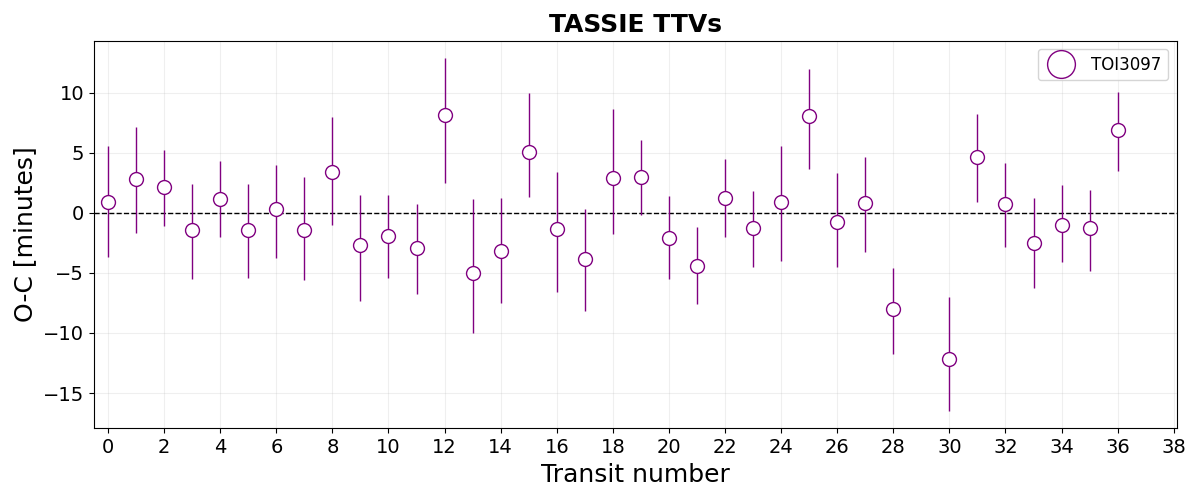}
	\end{subfigure}
	~
	\begin{subfigure}
		\centering
		\includegraphics[width=0.96\linewidth]{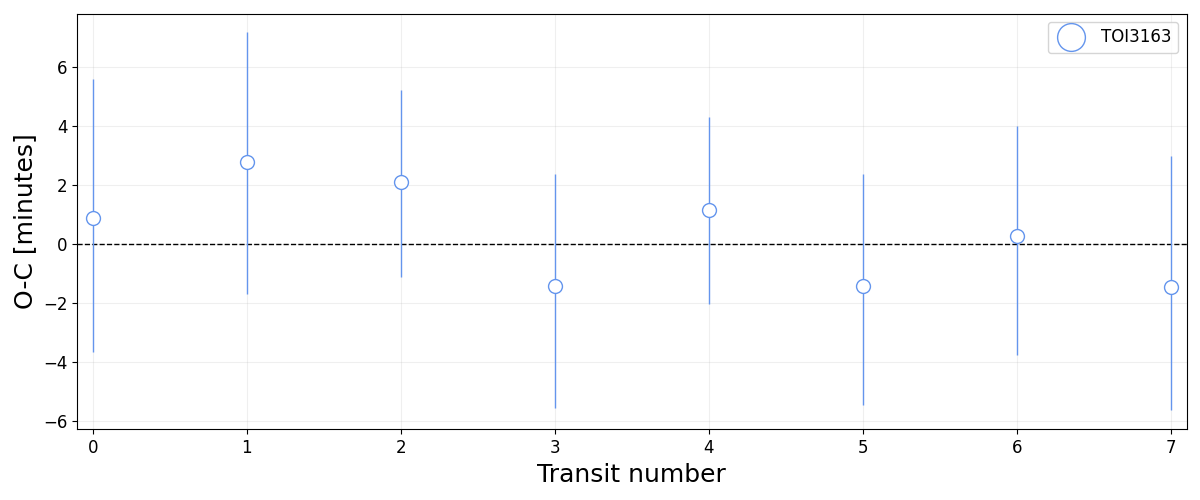}
	\end{subfigure}
 	\caption{Residuals between the observed and expected transit times for TOI3097Ab (top) and TOI3163b (bottom). The error bars correspond to the 1$\sigma$ uncertainty.}
	\label{fig:ttvs}
\end{figure}

The presence of undetected transiting or non-transiting planets in a system can act to perturb the expected Keplerian orbits of known transiting planets, resulting in TTVs and TDVs \citep{agolfabttvs}. We performed a search for TTVs for TOI3097Ab and TOI3163b using Juliet. We retained the best-fit parameters from the normal modelling run, this time passing priors for each orbital epoch. These were centred upon the values expected assuming a linear transit ephemeris. We then plot the residual between the observed and expected (calculated) transit mid times, as shown in Figure \ref{fig:ttvs}. For TOI3097Ab, we see minor deviations from expectations. However, these are not statistically significant, due to the large uncertainties ($\approx$ 4 minutes). This is validated by a significantly lower log-likelihood for the model ($\Delta ln(Z) = -454.21$). For TOI3163, we also find no evidence for TTVs above 2 minutes. 

\subsection{Photometric Performance}
Using the transit light curves, we can assess the quality of our data and photometry pipeline. We calculate the standard deviation of the raw residuals (without subtraction of the GP models) and measure the median photometric error for both targets. For TOI3097 (at r' = 13.48 $\pm$ 0.02), we find $\sigma_{3097}$ = 2.6 ppt and <$\delta F_{3097}$> = 3.2 ppt with exposure times of 90 seconds. For TOI3163 (at r' = 12.19 $\pm$ 0.03), we obtain $\sigma_{3163}$ = 1.8 ppt and <$\delta F_{3163}$> = 1.9 ppt with exposures of 120 seconds. Using our aforementioned ETC, we calculate the theoretical photon noise to be $\approx$ 1.5 ppt and $\approx$ 0.8 ppt for TOI3097 and TOI3163, respectively. This indicates that other sources of error are limiting the photometry, most likely being related to flat-field errors, atmospheric transparency changes and scintillation. Despite this, these values are consistent with other small telescope programs \citep[e.g.,][]{trappist_ppt, Minerva} and are sufficient for TASSIE's goals.
\newline\newline
To determine if the assumption of white noise is valid, we also test for correlations within the data. We perform a Pearson-R test on the residuals with various variables, such as time/phase, airmass, position on the CCD, sky background and FWHM. For TOI3097, all R-values are below $\pm$0.1 with the largest at R = 0.06 for correlation with FWHM. Furthermore, testing the null hypothesis of uncorrelated and normally distributed samples returns p-values all greater than 0.5. This indicates that the data is consistent with white noise. However, the same is not true for TOI3163 with weak correlations between almost all variables and the residuals (R = 0.2 - 0.4). This could be explained by two factors. Firstly, the TESS data shows more intrinsic scatter for TOI3163, implying stellar activity is adding red (correlated) noise. Secondly, the worse conditions during the TOI3163 observing session (reported in Section 3.1) may have also contributed to these correlations. This highlights that whilst our telescope and photometry pipeline is performing well generally, data quality will be situational and this may restrict our effectiveness in the follow-up of small planets. It also reiterates that careful detrending of light curves is important, validating the use of GPs in our transit modelling. 

\section{Conclusion}
The TESS mission has provided a large sample of new giant planet candidates to explore. We have presented the observations and analysis of two likely exoplanetary systems residing in the southern sky. The TOI3097 binary system hosts a sub-Jovian to Jovian size planet ($R_{3097Ab}$ = 0.89 $\pm$ 0.04 $R_{J}$) orbiting the primary K-dwarf star with a period of $P_{3097Ab}$ = 1.368386 $\pm$ 0.000006 days. The potentially low metallicity ([Fe/H] $\approx$ -1) of the system and proximity of the planet to the sub-Jovian desert makes it an exciting target for further follow-up studies. The TOI3163 system was found to contain a hot, Jupiter-size companion ($R_{3163b}$ = 1.42 $\pm$ 0.05 $R_{J}$ and $P_{3163b}$ = 3.074966 $\pm$ 0.000022 days) around a late F-type dwarf star in the thin disk. Radial velocity observations will be required to confirm these candidates are in the planetary mass regime. We also report three targets (TOI3070, TOI3124 and TOI4266) that we determine to be false positives, based upon their high RUWEs, statistical testing with TRICERATOPS and chromaticity. In future, we hope to compliment our photometric capabilities with high-resolution spectroscopic measurements. By searching for more elusive planets in the sub-Jovian desert, we will contribute to understanding the processes shaping the formation and evolution of hot, giant planets.

\begin{acknowledgement}
The authors wish to thank Caisey Harlingten for his generous support of the Greenhill Observatory. Without his donations, we would not have the H50 and 1.3 m telescopes. We thank Dr. Kym Hill, Mr. Keith Bolton, Dr. Tony Sprent and Dr. David Warren for their assistance in setting up the H50. We also thank the anonymous referee for their comments that helped to improve the manuscript.  We acknowledge J. B. Marquette for his preliminary ideas for the data reduction pipeline. 

This work uses data acquired at the Siding Spring Observatory with the Australian National University 2.3m Telescope. We acknowledge the traditional custodians of the land on which the telescope stands, the Gamilaraay people, and pay our respects to elders past and present. 

This paper includes data collected by the TESS mission, which are
publicly available from the Mikulski Archive for Space Telescopes
(MAST). Funding for the TESS mission is provided by NASA’s
Science Mission directorate. We acknowledge the use of public
TESS Alert data from pipelines at the TESS Science Office and at
the TESS SPOC. Resources supporting this work were provided
by the NASA High-End Computing (HEC) Programme through
the NASA Advanced Supercomputing (NAS) Division at Ames
Research Centre for the production of the SPOC data products. 

This work has made use of data from the European Space Agency
(ESA) mission \gaia (\url{https://www.cosmos.esa.int/gaia}), processed
by the Gaia Data Processing and Analysis Consortium (DPAC,  \url{https:
//www.cosmos.esa.int/web/gaia/dpac/consortium}). Funding for the
DPAC has been provided by national institutions, in particular
the institutions participating in the \gaia Multilateral Agreement. 

This research has made use of the SIMBAD data base and VizieR
catalogue access tool, operated at CDS, Strasbourg, France. 

This research has made use of NASA’s Astrophysics Data System. This research has made use of the NASA Exoplanet Archive, which is
operated by the California Institute of Technology, under contract
with the National Aeronautics and Space Administration under the
Exoplanet Exploration Programme. 

\end{acknowledgement}

\paragraph{Funding Statement}

This research is supported by an Australian Government Research Training Program (RTP) Scholarship.

\paragraph{Competing Interests}
The authors are not aware of any competing interests.

\paragraph{Data Availability Statement}
The code/notebooks used for analysis can be found on Github here: \url{https://github.com/tjplunkett/TASSIE-I}. Light curves and spectra will be uploaded to the TESS ExoFOP website after publishing. 

%\endnote in some journals will behave like \footnote; and \printendnotes will not output anything. 
%\printendnotes

%\bibliographystyle{yahapj}
\bibliography{refs}

\appendix
\section{Subdwarf Radius-Colour Relation}
\begin{figure}[h!]
    \centering
    \includegraphics[width=0.99\linewidth, height=0.8\linewidth]{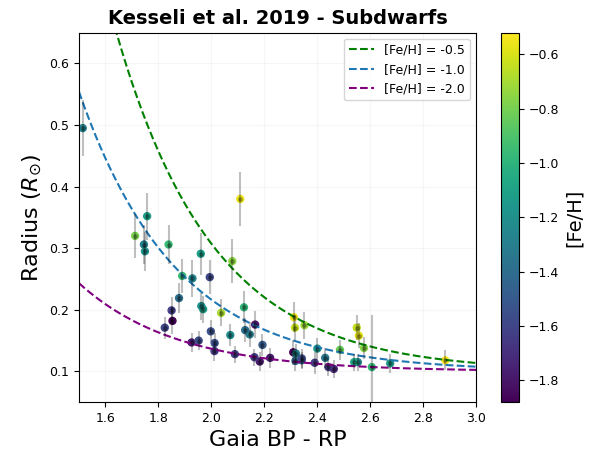}
    \caption{Stellar radius versus \gaia BP-RP colour for subdwarf stars from the Kesseli et al. (2019) sample. Data is coloured based upon the metallicity. The dashed lines show the best fit model at [Fe/H] = -0.5, -1.0 and -2.0.}
    \label{fig:kesseli_subdwarfs}
\end{figure}

Whilst trying to fit the radius for the potential sub-dwarf TOI3097B using the work of \citet{kesseli2019}, we noticed that their quoted relation (Eq. 5 in the paper) did not match the graphed results. The authors argued for a decreasing exponential form, requiring that no radius should be below $\approx$ 0.1 $R_{\odot}$ (due to degeneracy pressure). However, the form and coefficients listed did allow for such values. We therefore decided to re-fit the relations, slightly adjusting the form to:
\begin{equation}
    \frac{R}{R_{\odot}} = 0.1 + A \times e^{-(b(BP-RP) \, + \, c[Fe/H])}
\end{equation}
where, BP and RP are the \gaia blue and red bands magnitudes, respectively. We queried Vizier for their catalogue of 88 subdwarfs with photometry and derived stellar parameters. We cut any star with [Fe/H] > -0.5 dex to ensure the relations captured only subdwarf behaviour. Fitting the above relation, we find best fit coefficients of A = 84.9138 $\pm$ 36.1258, b = 2.7186 $\pm$ 0.1940 and c = -1.1537 $\pm$ 0.1341. The data and best-fit models for three metallicity bins are shown in Figure \ref{fig:kesseli_subdwarfs}.

\section{Auxillary Plots and Tables}
\begin{figure}[h!]
    \centering
	\includegraphics[width = 0.95\linewidth, height = 0.85\linewidth]{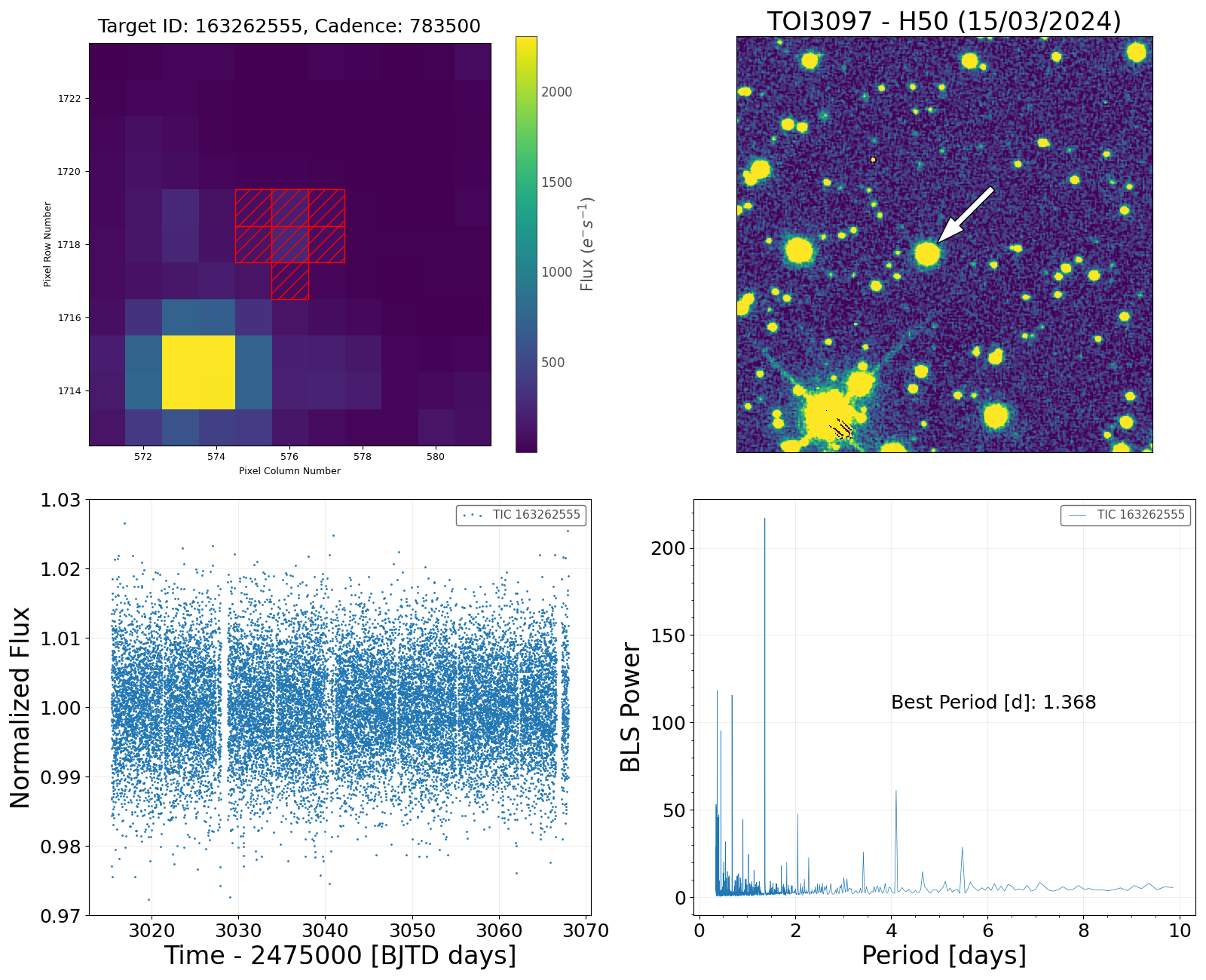}
	\caption{Summary plots for observations of TOI3097. The TESS Sector 64 TPF, normalized light curve and periodogram plots are shown. A cutout of the stacked H50 image is also displayed on the top right, matching the approximate orientation and scale of the TPF.}
	\label{fig:tpflcprd3097}
\end{figure}

\newpage 

\begin{figure*}
    \centering
    \includegraphics[width=1\linewidth]{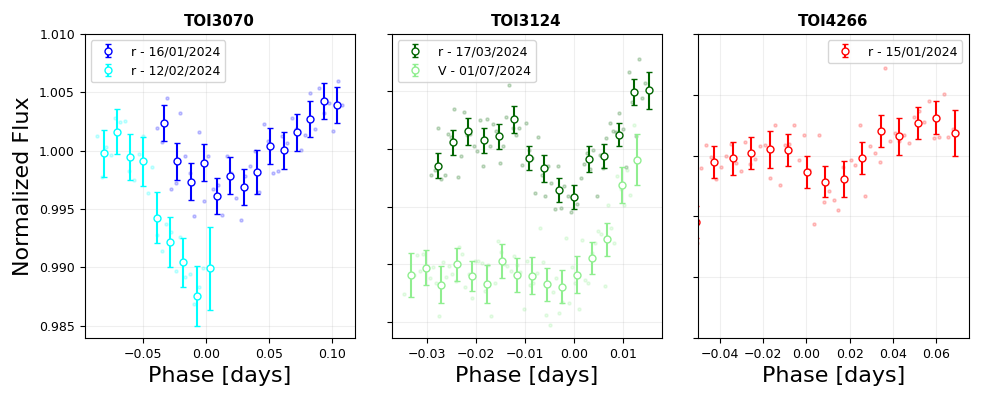}
    \caption{Light curves for non-planetary TOIs from the H50 telescope. Points with errorbars represent the data binned to 600 seconds.}
    \label{fig:nonplanets}
\end{figure*}

\begin{table*}[h!]
	\begin{center}
		\begin{tabular}{|p{2.5cm}|p{4cm}|p{2.5cm}|p{2.5cm}|p{2.5cm}|} 
			\toprule
			\headrow \textbf{Parameters} & \textbf{Description} & \textbf{TOI3097} & \textbf{TOI3124} & \textbf{TOI3163}\\
            \midrule
			\headrow \multicolumn{5}{|c|}{\textbf{Planetary Priors}} \\
            \midrule
			$P_{b}$ [days] & Orbital period & \textit{N}(1.368, 0.100) & \textit{N}(2.25, 0.100) & \textit{N}(3.072, 0.100) \\ 
            $T_{0}$ [BJD-2457000] & Time of transit centre & \textit{N}(3015.5, 0.05) & \textit{N}(3107.45, 0.05) & \textit{N}(3043.6, 0.05) \\ 
            $r_{1}$ & See Espinoza (2018) & \textit{U}(0,1) & \textit{U}(0,1) & \textit{U}(0,1) \\ 
            $r_{2}$ & " " & \textit{U}(0,1) & \textit{U}(0,1) & \textit{U}(0,1) \\ 
            $\sqrt{e} sin(\omega)$ & Parameterization for eccentricity and argument of periastron  & \textit{U}(-1,1) & \textit{U}(-1,1) & \textit{U}(-1,1) \\
            $\sqrt{e} cos(\omega)$ & " " & \textit{U}(-1,1) & \textit{U}(-1,1) & \textit{U}(-1,1) \\
            \midrule
			\headrow \multicolumn{5}{|c|}{\textbf{Instrumental Priors}} \\
            \midrule
            $D_{TESS}$ & Flux dilution factor for TESS & 1.0 & 1.0 & 1.0 \\
            $M_{TESS}$ & The offset relative flux for TESS & \textit{N(0, 0.1)} & \textit{N(0, 0.1)} & \textit{N(0, 0.1)} \\
            $J_{TESS}$ [ppm] & Photometric jitter for TESS & \textit{logU(0.1, 1000)} & \textit{logU(0.1, 1000)} & \textit{logU(0.1, 1000)} \\
            $q_{1, TESS}$ & Limb darkening coeff. for TESS &\textit{N(0.34, 0.03)} & \textit{N(0.32, 0.03)} & \textit{N(0.23, 0.02)} \\
            $q_{2, TESS}$ & "" & \textit{N(0.40, 0.04)} & \textit{(0.39, 0.04)} & \textit{N(0.36, 0.04)}\\
            $D_{H50}$ & Flux dilution factor for H50 & \textit{N(0.80, 0.01)} & 1.0 & \textit{N(0.94, 0.01)} \\
            $M_{H50}$ & The offset relative flux for H50 &\textit{N(0, 0.1)} & \textit{N(0, 0.1)} & \textit{N(0, 0.1)} \\
            $J_{H50}$ [ppm] & Photometric jitter for H50 &\textit{logU(0.1, 1000)} & \textit{logU(0.1, 1000)} & \textit{logU(0.1, 1000)}\\
            $q_{1, H50}$ & Limb darkening coeff. for H50 & \textit{N(0.47, 0.05)} & \textit{N(0.44, 0.04)} & \textit{N(0.32, 0.03)}\\
            $q_{2, H50}$ & " " &\textit{N(0.43, 0.04)} & \textit{N(0.42, 0.04)} & \textit{N(0.38, 0.04)} \\
            \midrule
			\headrow \multicolumn{5}{|c|}{\textbf{GP Priors}} \\
            \midrule
            $\sigma_{GP, TESS}$ [ppm] & Amplitude of the GP for TESS & \textit{logU($10^{-6}$, $10^{+6}$)} & \textit{logU($10^{-6}$, $10^{+6}$)} & \textit{logU($10^{-6}$, $10^{+6}$)}\\
            $\rho_{GP, TESS}$ & Timescale of the GP for TESS & \textit{logU($10^{-3}$, $10^{+3}$)} & \textit{logU($10^{-3}$, $10^{+3}$)} & \textit{logU($10^{-3}$, $10^{+3}$)}\\
            $\sigma_{GP, H50}$ [ppm] & Amplitude of the GP for H50 & \textit{logU($10^{-6}$, $10^{+6}$)} & \textit{logU($10^{-6}$, $10^{+6}$)} & \textit{logU($10^{-6}$, $10^{+6}$)} \\
            $\rho_{GP, H50}$ & Timescale of the GP for H50 & \textit{logU($10^{-3}$, $10^{+3}$)} & \textit{logU($10^{-3}$, $10^{+3}$)} & \textit{logU($10^{-3}$, $10^{+3}$)} \\
			\bottomrule 
		\end{tabular}
		\caption{The priors used in modelling of TASSIE TOIs with Juliet. Note: N, U and logU correspond to normal, uniform and log-uniform distributions.}
		\label{table:priors}
	\end{center}
\end{table*}

\begin{figure*}
    \centering
    \begin{subfigure}
    \centering
    \includegraphics[width=0.6\linewidth]{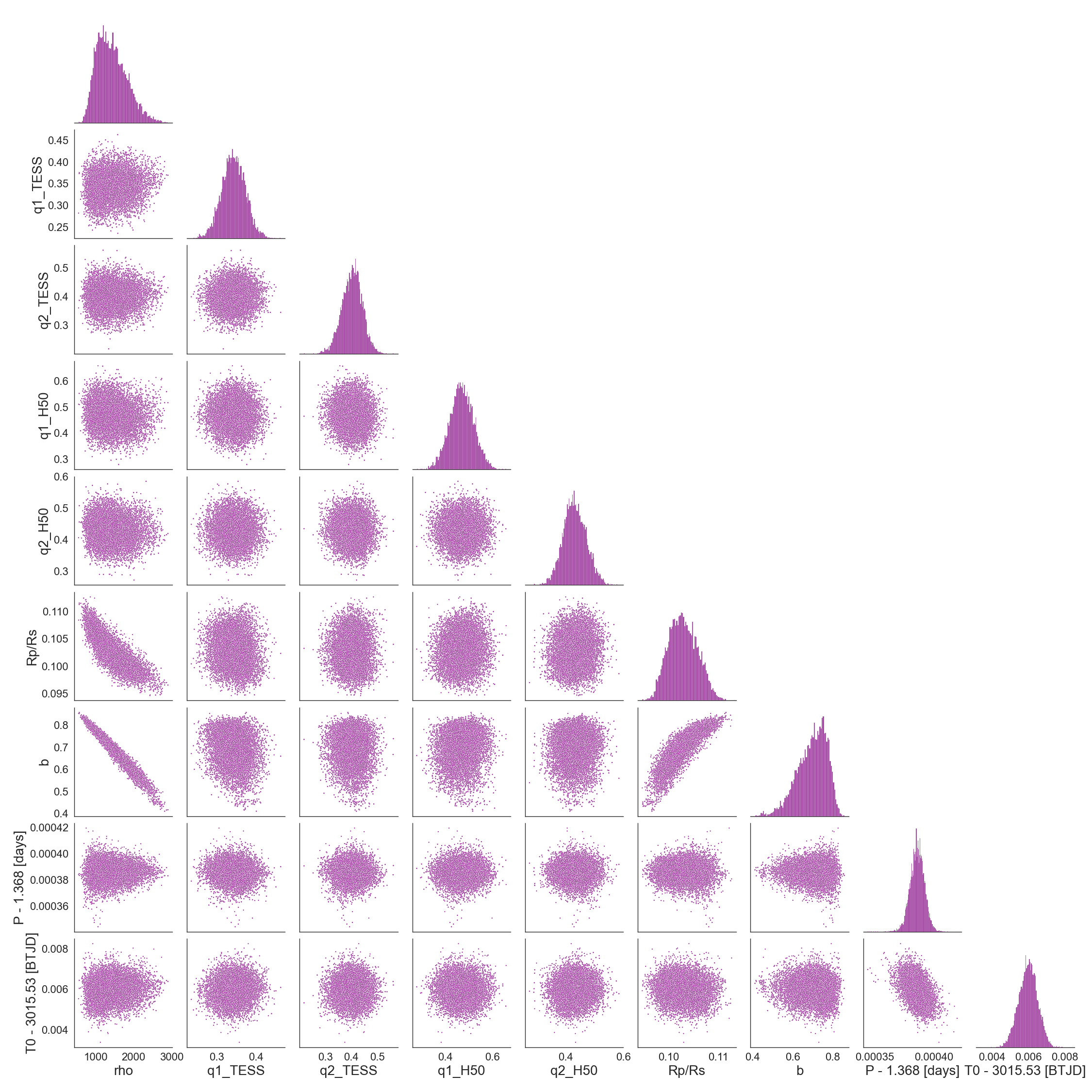}
    \end{subfigure}
    ~
    \begin{subfigure}
    \centering
    \includegraphics[width=0.6\linewidth]{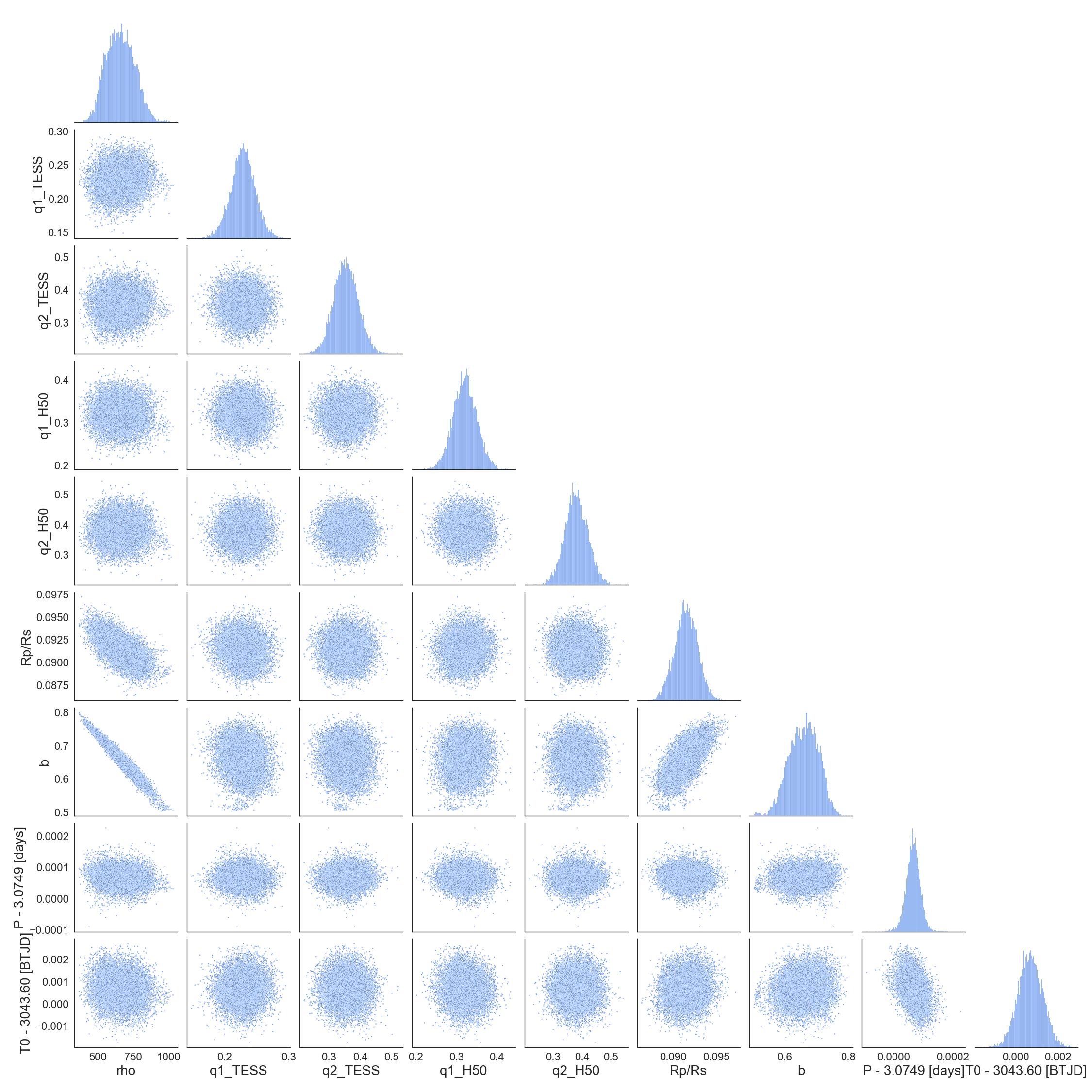}
    \end{subfigure}
    \caption{Corner plots of the posterior distributions from the modelling of TOI3097Ab (top) and TOI3163 (bottom) with Juliet.}
    \label{fig:posteriors}
    
\end{figure*}

\end{document}